\pdfoutput=1

\documentclass[%
reprint,
amsmath,amssymb,
aps,
prx,
]{revtex4-2}

\usepackage{graphicx}
\usepackage{picture}
\usepackage{placeins}
\usepackage{float}
\DeclareGraphicsExtensions{.pdf,.eps} 
\usepackage{amsmath,scalefnt}
\usepackage{mathtools}
\usepackage{amssymb}
\usepackage{verbatim}
\usepackage{amsmath,amsfonts,bbm}
\usepackage{amsbsy}
\usepackage{color}
\usepackage{cancel}
\usepackage{soul}
\usepackage{dsfont}
\usepackage[svgnames]{xcolor}

\DeclareMathOperator{\csch}{csch}

\usepackage[breaklinks=true,colorlinks=true,linkcolor=DarkBlue,urlcolor=DarkBlue,citecolor=DarkBlue]{hyperref}

\usepackage{subfigure}

\usepackage{braket}

\newcommand{\da}[0]{^{\dagger}}

\def\bs#1{\boldsymbol{#1}}
\def\ba#1{\left(\begin{array}{#1}}
	\def\ea{\end{array}\right)}
\def\bsm{\left(\begin{smallmatrix}}
	\def\esm{\end{smallmatrix}\right)}
\def\unit#1{\, \mathrm{#1}}

\def\bs#1{\boldsymbol{#1}}

\begin{document}

\title{Quantum Frequency Interferometry: \\with applications ranging from gravitational wave detection to dark matter searches}

\author{Richard Howl}
  \affiliation{QICI Quantum Information and Computation Initiative, Department of Computer Science, The University of Hong Kong, Pokfulam Road,
	Hong Kong}
	\affiliation{Quantum Group, Department of Computer Science, University of Oxford, Wolfson Building, Parks Road, Oxford, OX1 3QD, United Kingdom}
	\affiliation{School of Mathematical Sciences, University of Nottingham, University Park, Nottingham NG7 2RD, United Kingdom}

\author{Ivette Fuentes}
\email[Author to whom correspondence should be addressed: ]{I.Fuentes-Guridi@soton.ac.uk}
  \affiliation{School of Physics and Astronomy,
University of Southampton,
Southampton SO17 1BJ,
United Kingdom}
\affiliation{School of Mathematical Sciences, University of Nottingham, University Park, Nottingham NG7 2RD, United Kingdom}

\begin{abstract} 
We introduce a quantum interferometric scheme that uses states that are sharp in frequency and delocalized in position. The states are frequency modes of a quantum field that is trapped at all times in a finite volume potential, such as a small box potential. This allows for significant miniaturization of interferometric devices. Since the modes are in contact at all times, it is possible to estimate physical parameters of global multi-mode channels. As an example, we introduce a three-mode scheme and calculate  precision bounds in the estimation of parameters of two-mode Gaussian channels. This scheme can be implemented in several systems, including superconducting circuits, cavity-QED and cold atoms. We consider a concrete implementation using the ground state and two phononic modes of a trapped Bose-Einstein condensate.  We apply this to show that frequency interferometry can improve the sensitivity of phononic gravitational waves detectors by several orders of magnitude,  even in the case that squeezing is much smaller than assumed previously and that the system suffers from short phononic lifetimes. Other applications range from magnetometry, gravimetry and gradiometry to  dark matter/energy searches. 
\end{abstract}

\maketitle

Interferometers have become a powerful tool for precision measurements, often achieving sensitivities that are not possible using any other known technique. In the most common implementation, waves travel along two different spatial paths and are recombined creating an interference pattern. We will call this setup a {\it spatial interferometer} since the system follows two different trajectories in space. The phase difference along the paths is a $U(1)$ channel that encodes physical parameters, such as frequency and field strengths, that can be measured with very high precision. Interferometers can operate in both the classical and quantum regime.  The most remarkable application of a spatial interferometer is perhaps LIGO, which uses a very similar interferometer to that designed by Michelson to observe gravitational waves \cite{GWDetection}. Early measurements used classical light but recently the sensitivity has been enhanced by injecting quantum states of light through the output Faraday isolator \cite{Yu2020}. 

Interferometry in the quantum regime not only uses photons but also the wave character of massive particles, such as electrons, neutrons, atoms and molecules. Here, each particle generically follows a superposition of two spatial trajectories and interferes with itself when the paths recombine. The phase difference acquired by a spatial quantum superposition has been used, for example, to measure accelerations \cite{Peters_1999}; fundamental constants \cite{Rosi_2014}, including the fine-structure constant \cite{Parker_2018}; set constraints on dark energy models \cite{Jaffe_2017}; and has also been proposed as a method to detect gravitational waves at low frequencies \cite{savas}. For a review on the state of the art see \cite{bongs2019}.

 In spatial interferometry, cutting-edge sensitivity is usually limited by the available time of flight. This generically requires large spatial separations or long interferometer arms, which  can be several metres or kilometers long. Typically, spatial interferometers cannot be reduced in size without loosing precision. However,  the time of flight in atom interferometry can be increased by using Bragg diffraction and Bloch oscillations to slow down the particles  \cite{PhysRevA.94.043608}. In such schemes, interactions are undesirable because they reduce the coherence time of the interferometer \cite{Pereira:2017qyi}.  Here, we propose using interactions as a way to miniaturize devices while keeping high precision. The idea is to use interferometry in the frequency domain, or equivalently, temporal domain, and we refer to this as {\it quantum frequency interferometry}. In this case, waves do not follow different spatial paths but are quantum modes of vibration, such as those produced by interacting particles trapped in a localized potential. Since the modes are non-local and they can interact at all times, it is also possible not only to estimate quantities encoded in phase channels, but also to estimate the parameters of global unitaries, including entangling operations, in the Hilbert space of the modes.
 
 This new type of quantum interferometry is inspired by recent non-interferometric studies that consider estimating physical parameters using phonons of Bose-Einstein condensates (BECs). For example, frequency modes were used in a quantum metrology scheme to measure accelerations \cite{Accelerometer} and detect high-frequency gravitational waves using a BEC \cite{GWDetectorFirst,GWDetectorThermal}. More recently, phonons have been used in a method to miniaturize gravimeters \cite{Bravo2020b} and to measure the gravitational field gradient within the millimetre scale \cite{Bravo2020a}. In this paper, we extend these ideas to frequency interferometry, the application of which can improve by several orders of magnitude the sensitivities reported in these studies, which we illustrate  using the phononic gravitational wave detector \cite{GWDetectorFirst}.

\begin{figure}
\begin{subfigure}
  \centering
  \includegraphics[width=0.45\textwidth]{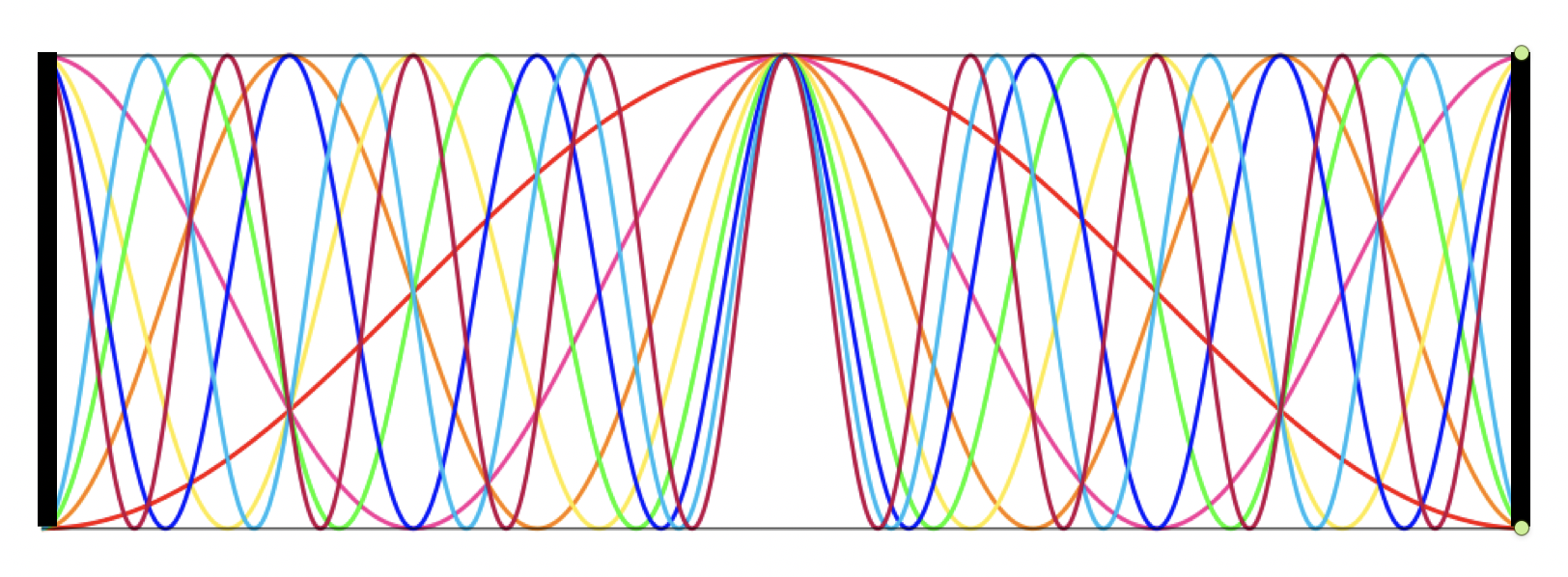}
  \label{fig:sfig0a}
\end{subfigure}%
\begin{subfigure}
  \centering
  \includegraphics[width=0.45\textwidth]{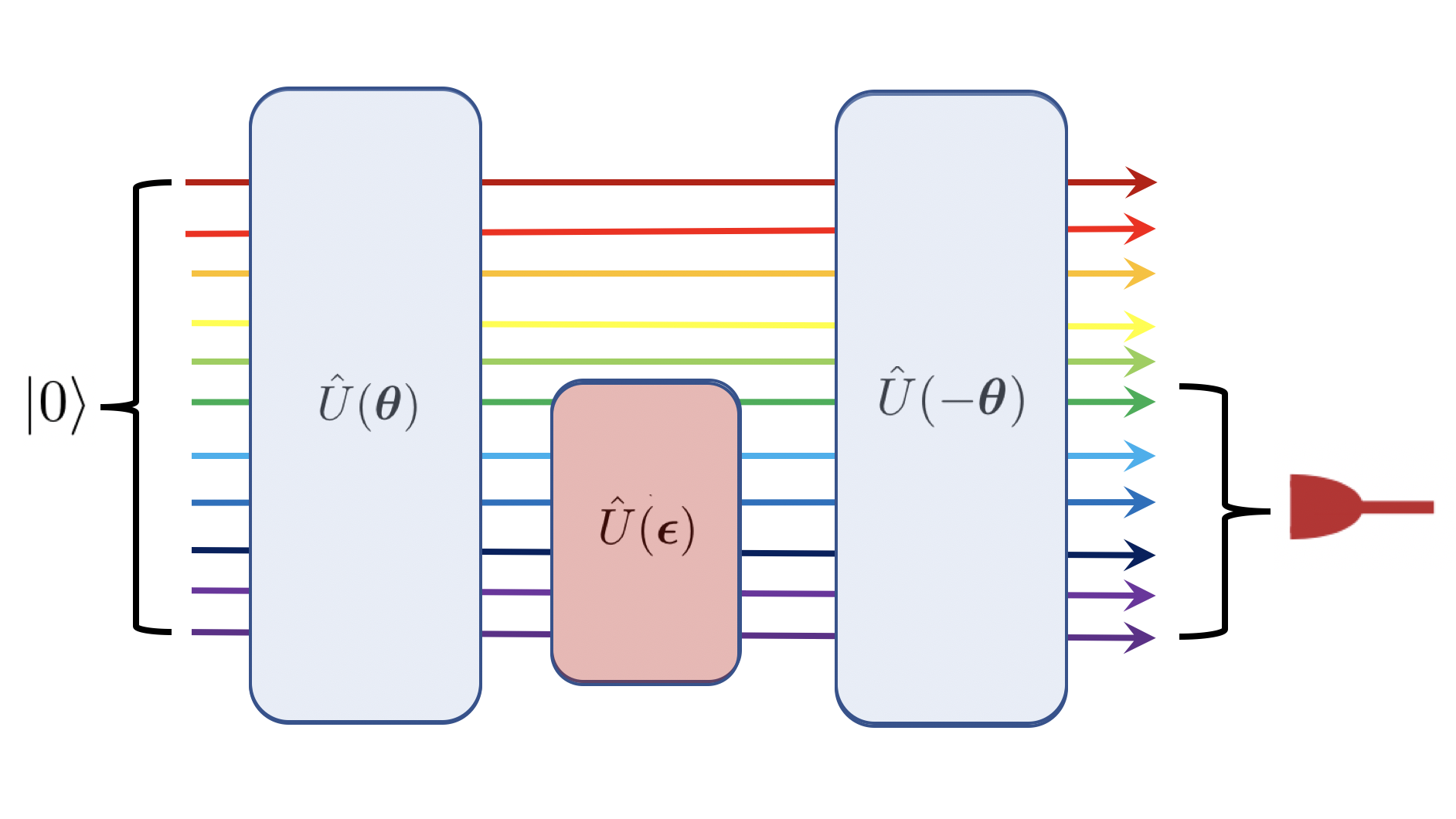}
  \label{fig:sfig0b}
\end{subfigure}
\caption{Quantum frequency  interferometry. Top: $N$ frequency modes of a box potential. Bottom: $N$ frequency modes, here taken to be in the vacuum, first pass through an active/passive beam splitter $\hat{U}(\boldsymbol{\theta})$. A subset then go through a unitary channel $\hat{U}(\boldsymbol{\epsilon})$  that imprints the parameters to be estimated. Finally, $\hat{U}(\boldsymbol{\epsilon})$ ``recombines'' the frequency modes.}
	\label{fig:frequencyG}
\end{figure}

\begin{figure}
\begin{subfigure}
  \centering
  \includegraphics[width=0.45\textwidth]{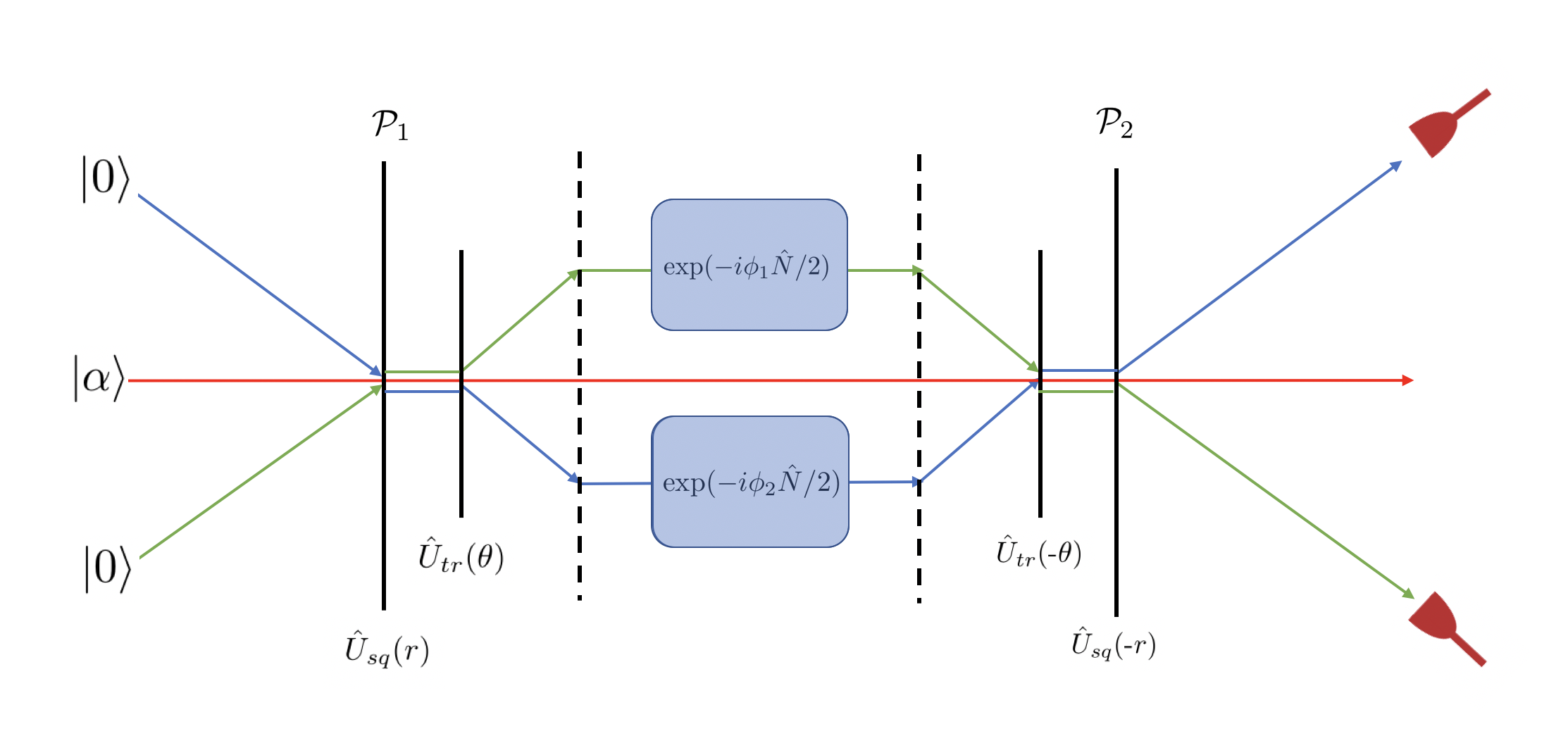}
  \label{fig:sfig1}
\end{subfigure}%
\begin{subfigure}
  \centering
  \includegraphics[width=0.45\textwidth]{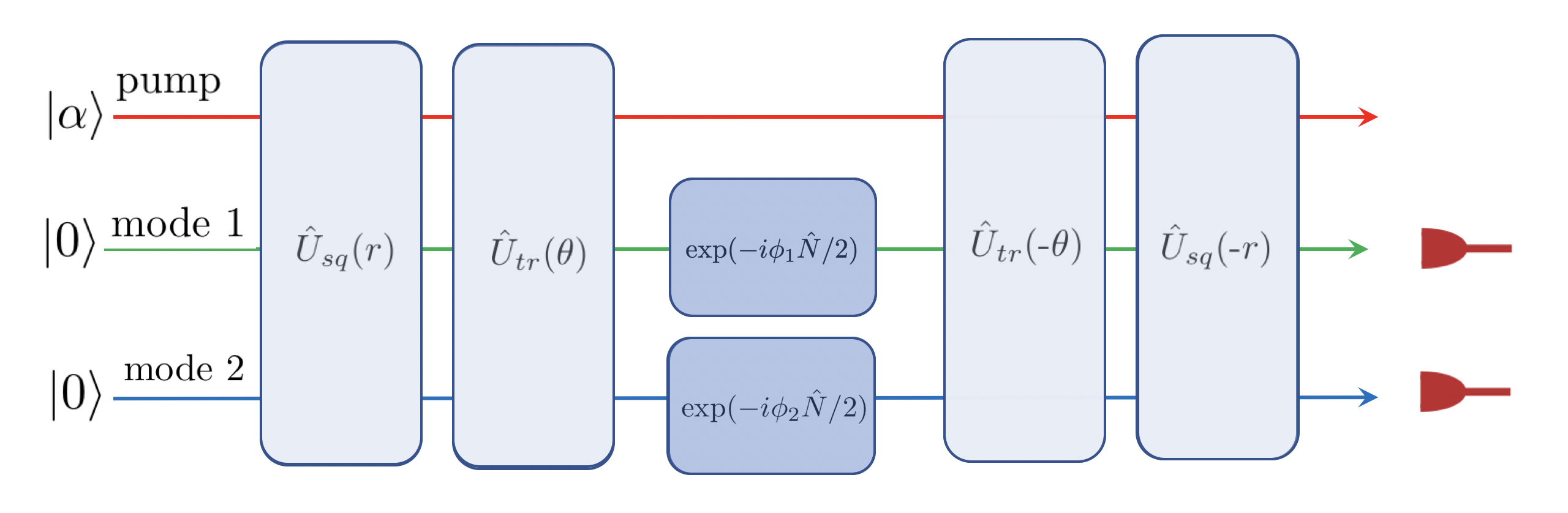}
  \label{fig:spatiala}
\end{subfigure}
\caption{Spatial interferometry. Top: pumped-up SU(1,1) scheme with phase channel \cite{PumpedUpSU11}. The pump mode (red) is initially in a coherent state and frequency modes (in blue and green) are in the vacuum state. $\hat{U}_{sq}(r)$ creates a two-mode squeezed state between the frequency modes. The side modes and pump are then sent through a tritter $\hat{U}_{tr} (\theta)$. The modes are separate and follow different trajectories in space. The side modes pick up a relative phase $\exp (-i (\phi_1 - \phi_2) \hat{N}/2)$. The dashed lines are laser pulses which act as atomic mirrors, or regular mirrors in the case of interferometry with photons. The modes are recombined through the reverse tritter and active beam splitter operations.  Finally, a number-sum measurement is performed on the side modes. Bottom: circuit representation of pumped-up SU(1,1) interferometry with phase channel.}
	\label{fig:spatial}
\end{figure}

\emph{Spatial interferometry and the pumped-up SU(1,1) scheme}. In general, a quantum interferometer can be broken up into four stages: i) the active/passive beam splitter, which usually entangles/populates  the input state, ii) the channel(s) that imprint(s) the parameter to be estimated, iii) the transformation that recombines the modes, facilitating   interference. Finally, in iv) the modes are measured. Conventionally, the second stage is considered to consist of unitaries acting independently in each spatial mode $\hat{U}(\phi) =U_{1}(\phi)\otimes U_2(\phi) \otimes \cdots$ (see also Figure \ref{fig:spatial}). 

Spatial interferometers, including the well-known Mach-Zehnder interferometer, are commonly SU(2) interferometers where  the states belong to a two-dimensional Hilbert space, which basis is spanned by the states corresponding to the two possible paths $\ket{1}$ and  $\ket{2}$ followed by the particles, and all the elements of the interferometer can be described in terms of quantum operators that obey an SU(2) algebra \cite{SU11}. For example, beam splitters exchange particles from one path to another and can be mathematically represented by $\sigma_{+}=\ket{1}\bra{2}$ and $\sigma_{-}=\ket{2}\bra{1}$. In 1986, Yurke et al.\ introduced a new type of interferometer where the passive beam splitters of SU(2), which perform mode-mixing operations, are replaced with active beam splitters, which perform quantum squeezing operations, such that the algebra that defines its operations is SU(1,1) \cite{SU11}.  This new type of interferometry, called SU(1,1) interferometry, has the advantage over SU(2) that squeezing and entanglement, which can improve the sensitivity of an interferometer, are generated within the interferometer itself rather than this having to be sourced separately.  Then, since the overall scheme will have fewer operational elements, it should, in principle, be more robust to noise. In this case, squeezing between spatial modes is introduced at points $\mathcal{P}_1$ and $\mathcal{P}_2$ in Figure \ref{fig:spatial}.  

Sending in a coherent pump to an SU(1,1) interferometer results in a two-mode squeezed state in the two spatial modes and, therefore, in principle, a $1/N$ scaling in sensitivity compared to $1/\sqrt{N}$ for a classical side mode state. However, generating a large number of particles in the spatial modes is extremely challenging and so the sensitivity is easily beaten by interferometers operating at the standard quantum limit with large input states. In order to overcome this issue, a variant of the SU(1,1) interferometer, called pumped-up SU(1,1), has recently been proposed where the pump beam first goes through a parametric amplifier and is then mixed with the side modes such that all particles take part in the estimation procedure \cite{PumpedUpSU11}. This essentially allows for a $1 / \sqrt{N N_0}$ scaling in sensitivity where $N_0$ is the number of particles in the pump beam and, in general, $N_0 \gg N$. This interferometer contains both passive and active beam-splitters, such that all the beam splitters within it are generated by general two and three-mode unitary Gaussian operations. In some sense this interferometer is  a generalisation of  SU(2) and SU(1,1) interferometry, with the parametric amplifiers of the SU(1,1) scheme seeding the quantum input state of an SU(2) (or more properly SU(3) since there are three modes) interferometer. Depending on the chosen three-mode mixing angle, it is also able to act like an SU(1,1) or SU(2) interferometer. After the three-mode mixing operation, the modes are physically separated and undergo unitary transformations that encode phases $\phi_1$ and $\phi_2$ on the respective states.  For example, in an SU(2) interferometer, the relative phase $\phi_1 - \phi_2$ is estimated, whereas, an SU(1,1) interferometer is  sensitive to the total unitary transformation  $\hat{U}(\phi) = \exp (-i \phi \hat{N}/2)$ where $\phi := \phi_1 + \phi_2$ and $\hat{N} := \hat{a}\da_1 \hat{a}_1 + \hat{a}\da_2 \hat{a}_2$, with $\hat{a}_1$ and $\hat{a}_2$ the annihilation operators for the two side modes \cite{SU11} (see Figure \ref{fig:spatial}). Note that the unitary can be written as two transformations acting independently on each mode $\hat{U}(\phi) = e^{-i \frac{\phi}{2} \hat{a}\da_1 \hat{a}_1} e^{-i \frac{\phi}{2} \hat{a}\da_2 \hat{a}_2}=U_{1}\otimes U_2$. 

\emph{Quantum frequency interferometry}. In matter-wave spatial interferometers, interactions are usually suppressed so that all particles are independent of each other. In that sense, they are single particle setups. The initial state is a product state of the individual particle states and the sensitivity scales with the standard quantum limit $1/\sqrt{N}$. However, quantum metrology studies show that entanglement, which requires interactions, can be used to increase precision, reaching, in the optimal case, the Heisenberg limit $1/N$. Interactions are key to frequency interferometry: the atomic interactions give rise to collective excitations, which are described by massless bosons (also known as phonons) that are used as the quantum information carriers.  

 We consider initially $N$ frequency phonon modes of a localized potential (Figure \ref{fig:frequencyG} Top). First the modes are prepared, often in the vacuum, then a passive or active beam splitter  $\hat{U}(\boldsymbol{\theta})$, with $\boldsymbol{\theta}=(\theta_1,\theta_2\dots\theta_N)$, is applied to create a quantum state of a subset of modes, which usually involves populating and entangling these modes. Subsequently, the modes (or an even smaller subset of modes) go through a channel $\hat{U}(\boldsymbol{\epsilon})$ with $\boldsymbol{\epsilon}=(\epsilon_1,\epsilon_2\dots\epsilon_m)$, which imprints the parameters to be estimated. Finally, before measurement, the modes are ``recombined'' by  $\hat{U}(-\boldsymbol{\theta})$.  The frequency modes of the phonons are delocalized in the potential but in physical contact with each other, allowing for global estimation channels where a global channel is a unitary that cannot be written as $U({{\epsilon}})\neq U_{1}\otimes U_{2} \otimes \cdots$. This, in particular, facilitates the generalization of pumped-up $SU(1,1)$ interferometry to global channels that include unitaries which entangle the modes.  Sensitivities are increased by preparing entangled states, which can be achieved using the particle interactions. The precision depends on the lifetime of the mode excitations, which can be extended by tuning the interactions without making the interferometer larger.  

\begin{figure}
\begin{subfigure}
  \centering
  \includegraphics[width=0.3\textwidth]{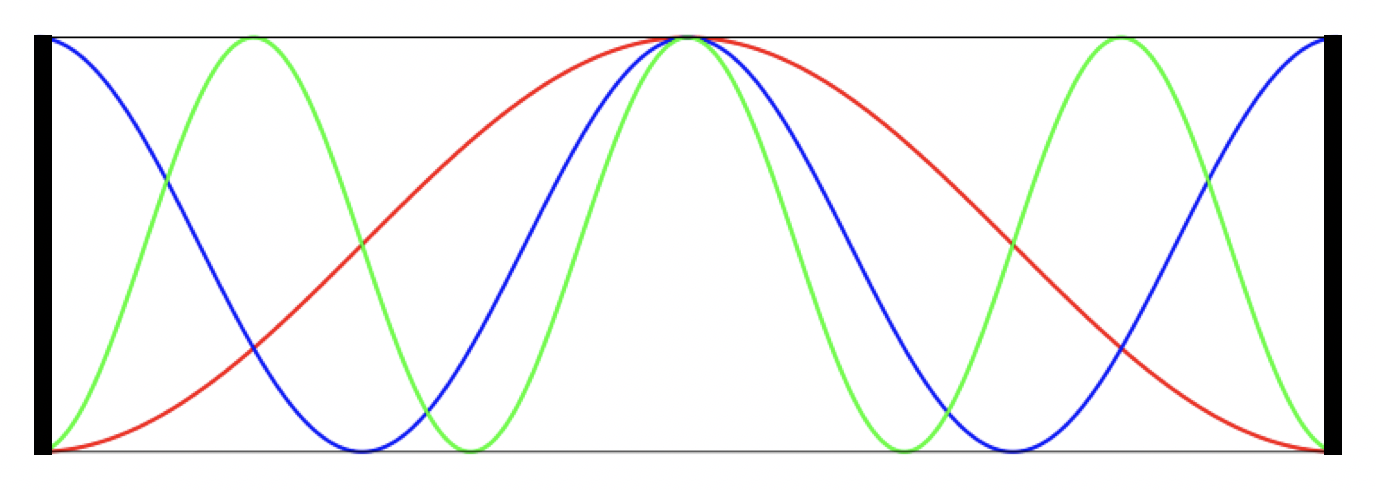}
 \end{subfigure}%
\begin{subfigure}
  \centering
  \includegraphics[width=0.45\textwidth]{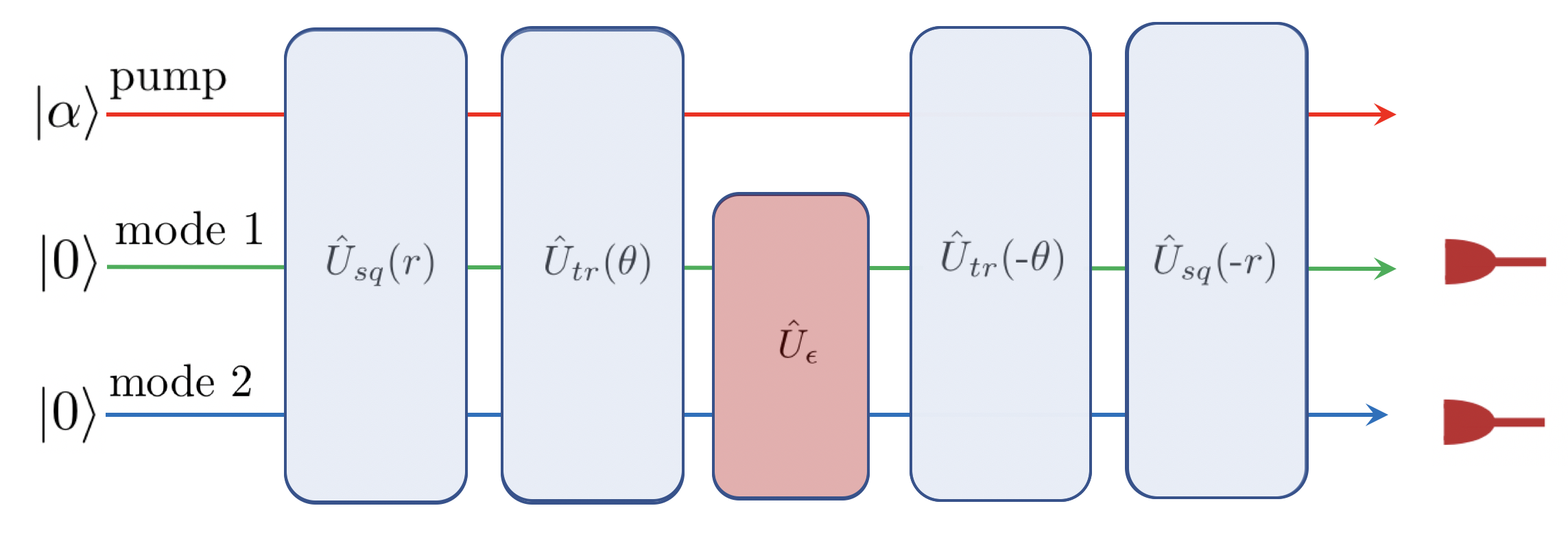}
\end{subfigure}
\caption{Three-mode example of Frequency Interferometry. Top: ground state and frequency modes in a box potential. The pump mode (red) is the phononic ground state and is initially in a coherent state. Frequency modes (green and blue) are sharp in frequency and delocalized in the potential. Two-mode squeezing $\hat{U}_{sq}(r)$ and mode mixing $\hat{U}_{tr} (\theta)$ can be implemented, for example, by periodically changing the box's length at the right frequency or by tailoring a given sequence of periodic motion at difference frequencies \cite{PhysRevApplied.10.044019,bruschi2016towards,PhysRevLett.111.090504,PhysRevD.86.105003,Bruschi_2013}. Bottom: Circuit representation of pumped-up SU(1,1) scheme with a Gaussian unitary channel. Since the modes are not spatially separated, it is possible to consider channels that act on the modes globally. In this case, the green and blue modes undergo a two-mode Gaussian unitary channel $\hat{U}_{\epsilon}$. After the reverse tritter and active beam splitter operations, a number-sum measurement is performed on the frequency modes.}
\label{fig:frequency}
\end{figure}

\emph{Frequency interferometry with Gaussian channels}. Here, we have chosen to present a three-mode example of frequency interferometry, illustrated  in Figure \ref{fig:frequency}. The example uses an analogue in frequency space of the squeezing and tritter operations used in the pumped-up SU(1,1) scheme. However, the pumped-up SU(1,1) was introduced for spatial interferometry and, therefore, restricted to separable channels of the form $\hat{U}(\phi)=U_{1}\otimes U_2$. In this example, the channels we consider are entangling two-mode Gaussian channels given by two-mode squeezing and mode-mixing channels,
\begin{align} \label{eq:Us}
U(\xi)&= e^{\xi \hat{a}_1\da \hat{a}_2\da - \xi^{\ast} \hat{a}_1 \hat{a}_2}~~\mathrm{or}\\ \label{eq:Um}
 U(\zeta)&= e^{\zeta \hat{a}_1\da \hat{a}_2 - \zeta^{\ast} \hat{a}_1 \hat{a}_2\da},
\end{align}  
respectively, where $\xi:= s e^{i \phi_{B}}$ and $\zeta := m e^{i \phi_{A}}$, with  $s \geq 0$, $m \geq 0$ and $\phi_A, \phi_B \in \mathbb{R}$. These channels include both SU(1,1) and SU(2)-like interferometry. Together with the phase-shift channel, considered in \cite{PumpedUpSU11} and single-mode squeezing, they form the complete set of unitary two-mode Gaussian channels, also known as a Bogoliubov transformations \cite{CMF}.  An analysis of the optimum Gaussian input states for Gaussian channels using the Quantum Fisher Information (QFI) is given in \cite{DomOptimum}.
 
We will consider the parameter of interest $\epsilon$ to be encoded in the squeezing parameter $s$ and mode-mixing parameter $m$. Specifically, we take  $s=: \frac{1}{4} \epsilon B$ and $m=: \frac{1}{4} \epsilon A$, where $A$ and $B$ depend on the physical quantities of the specific implementation of the scheme.  Since the input is a Gaussian state and all operations within the interferometer are Gaussian, we find it most straightforward to consider the operation of the new interferometer within the covariance matrix formalism (CMF) (see e.g.\ \cite{CMF}), where just two finite matrices are needed to describe the quantum states of the modes. This formalism allows for simple calculations of the QFI for the device, where we surprisingly find that the sensitivity can be slightly enhanced when the parameter of interest is encoded in a squeezing or mode-mixing channel.

 Although the QFI can provide the optimum precision given the above first two stages of an interferometer, it does not by itself identify the measurement scheme that achieves this.  However, we find that the sensitivity of a simple intensity measurement scheme can saturate the quantum Cram\'{e}r-Rao bound (QCRB) at large input  particle number. Within the CMF, we also find simple expressions for the sensitivity of general interferometers with such intensity measurement schemes, which can be straightforwardly generalised to other schemes such as homodyne.

The CMF  is a phase-space representation of a quantum state where a Gaussian state is fully defined by its displacement vector $\bs{d}$ and covariance matrix $\bs{\sigma}$.  In the real $q-p$ representation, these  are defined as the following  for a  system consisting of $n$ bosonic modes \footnote{Note that several conventions are used for the definitions of $\bs{d}, \bs{\sigma}$ and $\hat{\bs{x}}$. See, e.g.\ \cite{CMF}, for an alternative convention.}:
\begin{align}
\bs{d} &:= \braket{\hat{\bs{x}}},\\
\bs{\sigma}_{ij} &:= \frac{1}{2} \braket{\{\hat{\bs{x}}_i, \hat{\bs{x}}_j\}} -  \braket{\hat{\bs{x}}_i} \braket{\hat{\bs{x}}_j},  
\end{align}

where $\hat{\bs{x}} := (\hat{x}_1, \hat{x}_2, \ldots  \hat{x}_{2n-1}, \hat{x}_{2n})^T$ and $\hat{x}_i$ are quadratures defined by:
\begin{align}
\hat{x}_{2i - 1} &:=  \hat{a}_i + \hat{a}_i\da,\\
\hat{x}_{2i} &:= i (\hat{a}_i\da - \hat{a}_i),
\end{align}
with $i \in \mathds{Z}^+$, and $\hat{a}_i$ and $\hat{a}_i\da$ the annihilation and creation operators.  Unitary transformations $\bs{U}$ acting on density matrices now lead to symplectic matrices $\bs{S}$ acting on the displacement and covariance matrices through $\bs{d}' = \bs{S} \bs{d}$ and $\bs{\sigma}' = \bs{S} \bs{\sigma} \bs{S}^T$ \cite{CMF}. 

The initial state of the pump mode is assumed to be a coherent state. As in standard SU(1,1) interferometry, we act on this state with a two-mode squeezing operation $\hat{U}_{sq}(r) = \exp \{ \chi (\hat{a}_1\da \hat{a}_2\da - \hat{a}_1 \hat{a}_2)\}$ to parametrically populate the side modes, where $\chi := r \exp \{ i \vartheta_{sq} \}$. The state of the full system is then given by $\bs{S}_{s} \bs{d}_0$ and   $\bs{S}_{s} \bs{\sigma}_0 \bs{S}_{s} ^T$ where $\bs{S}_{s}$ is the symplectic matrix of this squeezing unitary and $\bs{d}_0$ and $\bs{\sigma}_0$ of the displacement and covariance matrices of the initial coherent state \footnote{See Appendix \ref{app:SympleticMatrices} for real $q,p$ representations of the symplectic matrices corresponding to all the unitary processes involved in the interferometer.}.   Here we have assumed that the pump is fairly undepleted by the squeezing operation and remains in a coherent state $|\alpha \rangle$, but we take $\alpha \rightarrow \alpha_0$ after acting with $\bs{S}_s$, where  $|\alpha_0^2|:= |\alpha|^2 - 2 \sinh^2 r$, and $N_0 := |\alpha_0|^2$, $N := 2 \sinh^2 r$, so that particle number is conserved \cite{PumpedUpSU11}.  

Next we apply a tritter to the three modes, whose symplectic matrix we denote by  $\bs{S}_{tr}$. We then act on the side modes with the squeezing or mode-mixing operations given by \eqref{eq:Us} and \eqref{eq:Um}. Subsequently, the beams are brought `back together' with another tritter and then an outcoupling process, which are both the reverse of the operations that were performed prior to the Gaussian unitary channel.  The state of the  full interferometer is then defined by $\bs{d} = \bs{S} \bs{d}_0$ and $\bs{\sigma} = \bs{S}  \bs{\sigma}_0 \bs{S}^T$, where $\bs{S}:=\bs{S}_{-} \bs{S}_{\epsilon} \bs{S}_+$ with $\bs{S}_{-} := \bs{S}_s (-r) \bs{S}_t (-\theta)$, $\bs{S}_+ :=  \bs{S}_t (\theta) \bs{S}_s(r)$ and $\bs{S}_{\epsilon}$ being either the squeezing or mode-mixing channel for the side modes. 

\begin{figure}
	\begin{center}
		\includegraphics[width=0.4\textwidth]{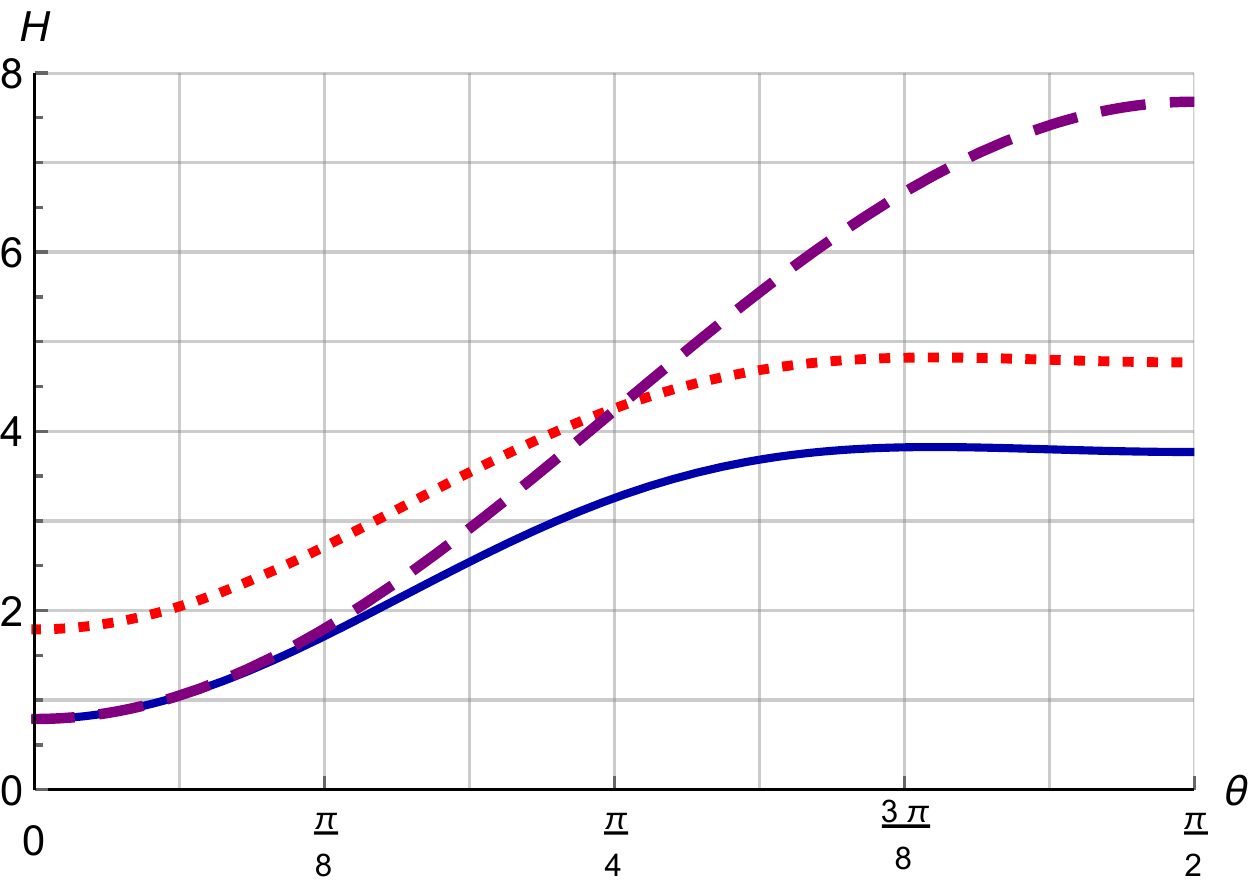}
	\end{center}
	\vspace{-0.5cm}
	\caption{The QFI, $H_{\epsilon}$, of the interferometer as a function of the tritter angle $\theta$, with $r=0.4$ and $|\alpha_0|^2 = 3.4$. The solid blue line is when a phase-shift channel is used, the dotted red line is for a two-mode squeezing channel  (with $\nu_B = \pi/2$, $\vartheta_{sq} = \phi_B + \pi/2$, $B=2$), and the dashed purple line is for a two-mode mode-mixing channel (with $\nu_A = \pi/2$, $\phi_A = 0$, and $A=2$).}
	\label{fig:QFIs} 
\end{figure}

\emph{Quantum Fisher information}. Since it is independent of the particular measurement scheme used, when calculating the QFI, we only need to consider the operations up to and including the Gaussian unitary channels i.e.\ the state of the relevant system is defined by $\bs{d} = \bs{S}_{\epsilon} \bs{S}_+ \bs{d}_0$ and $\bs{\sigma} = \bs{S}_{\epsilon} \bs{S}_+ \bs{d}_0  (\bs{S}_{\epsilon} \bs{S}_+)^T$.  For Gaussian states, the QFI, $H_{\epsilon}$, can be obtained through simple expressions (see e.g.\ \cite{QFIMatrices,QFIMatricesBraun}). When the squeezing channel is chosen, the QFI is:
	\begin{align} \nonumber
	H_{\epsilon} = \frac{1}{16} B^2 \Big[ 4 &+ \sin^2 (2 \theta) \sinh^2 r \\ \nonumber &+ 2 (1 + \cos^4 \theta) \eta_2(\vartheta_{sq}) \sinh^2 (2r) \\\label{eq:HSqFull} &+|\alpha_0|^2 \Big(4 \sin^4 \theta + \eta_1(r) \sin^2 2 \theta \Big)\Big],
	\end{align}
where:
\begin{align} \nonumber
\eta_1(r) &:= \sinh (2r) \cos \nu_{B} + \cosh (2r),\\\nonumber
\eta_2(\vartheta_{sq}) &:= \sin^2 (\vartheta_{sq} - \phi_{B}),\\\nonumber
\nu_{B} &:= 2 \vartheta - 2 \vartheta_0 - \vartheta_{sq} + 2 \phi_{B}.
\end{align}
 In Figure \ref{fig:QFIs}, we plot the dependence on the tritter angle $\theta$ for the QFI when there is a squeezing channel \eqref{eq:HSqFull}, standard phase-shift  channel (which can be found in \cite{PumpedUpSU11} and Appendix \ref{app:PhaseShift}) and mode-mixing channel \eqref{eq:HmFull}. Interestingly, we find a slight improvement in the QFI for the squeezing and mode-mixing channels compared to the conventional phase-shift channel. In the former case this is because even a vacuum input to the channel can be used to estimate the parameter of interest.

Given the optimum phase relationships $\vartheta_{sq} = \phi_B + \pi/2$ and $\nu_B = 0$, the QFI for the squeezing channel \eqref{eq:HSqFull} has turning points at $\theta = 0$, $\theta - \pi/2$ and $\theta = \theta_t$ where $\theta_t$  is defined in  \cite{PumpedUpSU11} and Appendix \ref{app:TurningPoints}, and approximates $\pi/4 + \csc^{-1} (N + \sqrt{N(N+2)})$ when $\overline{N}$ is large. Here we concentrate on the most interesting and relevant regimes, which are when  $\theta = 0$ and $\overline{N} \gg 1$. In the former case, we recover standard SU(1,1) interferometry, and so \eqref{eq:HSqFull} becomes the QFI for an SU(1,1) interferometer with a squeezing channel, which could still be considered an SU(1,1) interferometer since the unitary representation of a squeezing channel is part of the SU(1,1) group and the number-sum operation can still be used in the measurement scheme. The QFI in this case scales as $N^2$, which is the scaling obtained in conventional SU(1,1) interferometry. On the other hand, when $\overline{N} \gg 1$, the QFI \eqref{eq:HSqFull} can be approximated by $ B^2 \sin^2 (2 \theta) N \overline{N} / 8$, where  we have also assumed that $N \gg 2$ and taken  the optimum phase relation  $\nu_B = 0$. This $N \overline{N}$ scaling was also found in \cite{PumpedUpSU11} for the phase-shift channel case. In practice, this scaling can beat the conventional $N^2$ scaling by orders of magnitude since $\overline{N} \gg N$ in current experiments.

If   instead of the squeezing channel we use the mode-mixing channel, the QFI is:
\begin{align}  \nonumber
H_{\epsilon}&=\frac{1}{8} A^2 \Big[ (1 + \cos^2 \theta) \sinh^2 (2r) \\ \nonumber &+ \sin^2  \theta ~\Phi_1(\theta,\phi_A) \Big( \sinh^2 (2r) - 2 \sinh^2 r\Big)\\ \label{eq:HmFull} &+2 |\alpha_0|^2 \sin^2 \theta \Big(\sin^2 \theta \sin^2 \phi + \Phi_1 (\theta,\phi_A) \eta_3 (r)\Big) \Big],
\end{align}
where:
\begin{align}\nonumber
\Phi_1(\theta,\phi_A) &:= \sin ^2\theta   \sin^2 \phi_{A} -1,\\\nonumber
\eta_3(r) &:=  \sinh (2r) \cos \nu_A  -  \cosh (2r),\\\nonumber
\nu_A &:= 2 \vartheta - 2 \vartheta_0 + \vartheta_{sq}
\end{align}
 As for the squeezing channel, when $\theta = 0$ we obtain the QFI for a conventional SU(1,1) interferometer with a mode-mixing channel. In contrast to the squeezing case, however, this type of interferometer derived when $\theta = 0$ would not be considered an SU(1,1) interferometer by the original definition \cite{SU11} since the unitary representation of the mode-mixing channel does not form part of the SU(1,1) group. Instead, such an interferometer would be described by a larger group, for example,  the unitary group associated with a double covering of Sp($4, \mathbb{R}$) \cite{Arvind1995}. Similarly, our full interferometer with a general unitary Gaussian channel could be considered  an Sp($6, \mathbb{R}$) interferometer. 
 If we assume $\overline{N} \gg 1$ in \eqref{eq:HmFull},  we obtain $H \approx A^2 \sin^2 \theta (1-\sin^2 \theta \sin^2 \phi_A)\overline{N} N/2$, where  we have also assumed that $N \gg 1/ 2$ and  taken   $\nu_A = \pi$.

\emph{Sensitivity}.  We now choose a specific measurement process, the sum of the number of particles in the side modes. That is, the measured observable is $\hat{S} = \hat{N} := \hat{a}^{\dagger}_1 \hat{a}_1 + \hat{a}^{\dagger}_2 \hat{a}_2$. The square of the sensitivity of the interferometer is defined as (see e.g.\ \cite{QuantumLimitsOptics} for a derivation):
\begin{align} \label{eq:Sens}
\Delta^2 \epsilon := \frac{\mathrm{Var}(\hat{S})}{   (\partial_{\epsilon} \braket{\hat{S}})^2 },
\end{align}
where $\mathrm{Var}(\hat{S}) := \braket{\hat{S}^2} - \braket{\hat{S}}^2$. For Gaussian states, this is related to the Fisher information through \cite{GaussianActive}:
\begin{align}
F = F_0  + \frac{2 \Big(\partial_{\epsilon} \sqrt{\mathrm{Var}(\hat{S})}\Big)^2 }{\mathrm{Var}(\hat{S}) },
\end{align}
where $
F_0 := 1 / \Delta^2 \epsilon$, 
such that  $ F \geq F_0$.  Writing \eqref{eq:Sens} in the CMF in the $q,p$ basis, we find the simple expressions \footnote{See Appendix \ref{app:Heterodyne} for the equivalent expressions for homodyne/heterodyne measurements.}:
\begin{align} \label{eq:S}
\braket{\hat{S}} &= \frac{1}{4} [\mathrm{Tr}(\bs{\sigma}_s) + \bs{d}^T_s \bs{d}_s - 2n],\\ \label{eq:VarS}
\mathrm{Var}(\hat{S}) &= \frac{1}{8} [  \mathrm{Tr}(\bs{\sigma}^2_s) + 2 \bs{d}^T_s \bs{\sigma}_s \bs{d}_s - 2n] 
\end{align}
where $n$ is the number of modes, which is $2$ in this case, and $\bs{\sigma}_s$ and $\bs{d}_s$ are the covariance and displacement matrices of the side modes, which are generated from the full $\bs{d}$ and $\bs{\sigma}$  by simply removing the first two rows and columns. Expressions for the mean $\langle \hat{S} \rangle$ and variance $\mathrm{Var}(\hat{S})$ for the squeezing channel can be found in Appendix \ref{app:Sensitivity} when working with small $s$. In the limit that $\overline{N} \gg 1$ and $N \gg 1$,  we  find $F_0 = B^2 \sin^2 (2 \theta) N \overline{N} / 8$, 
where we have also taken  the optimum phase relation  $\nu_B = 0$.  This matches the corresponding QFI  and so the  number-sum measurement is an optimum measurement scheme  in these limits.

When working with small $m$, full expressions for the mean $\langle \hat{S} \rangle$ and variance $\mathrm{Var}(\hat{S})$ for the mode-mixing channel can be found in Appendix \ref{app:Sensitivity}. Taking $\overline{N} \gg 1$ and $N \gg 1$, we obtain $
F_0 \approx A^2 \sin^2 \theta (1-\sin^2 \theta \sin^2 \phi_A)\overline{N} N /2$, 
where we have also taken  the optimum phase relation  $\nu_A = \pi$. Therefore, as with the squeezing channel, this matches the  QFI  in the corresponding limits. Future work will apply multiparameter estimation techniques to, for example, estimate squeezing and mode-mixing parameters simultaneously.
 
\emph{Implementation in a BEC}. As an example of an implementation of the scheme, consider a one-dimensional BEC trapped in a box of length $L$ and interacting with a periodic potential $\epsilon \mathcal{V}_{\epsilon}(x,t)$, with $\epsilon \ll 1$. The Hamiltonian for the Bose gas is given by
\begin{align} \label{eq:fullH2}
\hat{H} =  &\int  dx \hat{\Psi}^{\dagger} \Big[-\frac{\hbar^2}{2m} \nabla^2 + \epsilon \mathcal{V}_{\epsilon}(x,t) \Big]\hat{\Psi} \\
&+ \frac{1}{2} g \int  dx \hat{\Psi}^{\dagger} \hat{\Psi}^{\dagger} \hat{\Psi} \hat{\Psi},
\end{align}
where $\hat{\Psi}$ is the field operator of the atoms.  The constant $g = 4 \pi \hbar^2 a / m$ is the coupling strength for a two-body contact  potential, with $a$  the s-wave scattering length and $m$ the atomic mass \cite{PitaevskiiBook,LandauLifshitzQM}. Depending on the nature of the potential $\epsilon \mathcal{V}_{\epsilon}(x,t)$, this implementation can be used to measure gravitational effects including gravitational fields and their gradient \cite{R_tzel_2018}, gravitational waves \cite{GWDetectorFirst} and also electromagnetic fields interacting with the phonons \cite{bravo_analog_2015,PhysRevD.98.025011,pJuschitz}.  

For a large number of atoms $N_0\gg 1$ in the dilute and low-energy regime, the Bogoliubov approximation enables the decomposition of the atomic field operator $\hat{\Psi} (x,t) =  [ \phi_0{a}_0 + \hat{\psi}(x,t) ] e^{- i \mu t / \hbar}$ in terms of a classical function for the ground state $\phi_0{a}_0\approx\phi_0\sqrt{N_0}$ and a quantum field $\hat{\psi}(x,t)$ that can be written as an infinite sum of quasi-particle modes,  with $\mu$  the chemical potential. At low energies, these quasi-particles behave as phonons with frequency $\omega_n=n\pi c_s/L$, where  $c_s$ is the speed of sound. In Appendix \ref{app:BEC}, we show how different choices of potentials $\mathcal{V}_{\epsilon}(x,t)$ and resonance conditions can implement $\bs{S}_{s}(r)$, $\bs{S}_{t}(\theta)$ and $\bs{S}_{\epsilon}$ for the phonons. 

Consider, for example, taking $\mathcal{V}_{\epsilon}(x,t) =  \mathcal{V}_0  x^2 \sin \Omega t$, with $\mathcal{V}_0$ and $\Omega$ constants. If two phonon modes $l$ and $n$ have frequencies such that $\Omega = \omega_n + \omega_l$, then this potential will generate a resonant two-mode squeezing Gaussian channel that has squeezing parameter $s=\frac{\epsilon B}{4}= 2 \epsilon |\mathcal{M}_{l n}\mathcal{V}_{l,n}|t/\hbar$, with:
\begin{align}
\mathcal{M}_{l n} & \approx \mathrm{i}  \frac{L^{3}\left(l^{2}+n^{2}\right)}{2 \sqrt{2 n l}\left(l^{2}-n^{2}\right)^{2} \pi^{3} \zeta}.
\end{align}
Here, $\zeta := \hbar / (\sqrt{2} m c_s)$ is the BEC healing length., and the coefficients  $\mathcal{V}_{l,n}$ are given by $\mathcal{V}_{l,n}=\left(1+(-1)^{l+n}\right)\mathcal{V}_{0}$ \cite{R_tzel_2018}. If we assume that $\bs{S}_{s}(r)$ creates $N_P=2\sinh^2 r$ phonons in modes $n$ and $l$, and that the squeezing channel is generated by a quadratic potential, the precision in estimating $\epsilon$ is then:
\begin{align}
\Delta \epsilon = \frac{\hbar^2 \pi^{3}\sqrt{2 n l}\left(l^{2}-n^{2}\right)^{2}}{m c_s\mathcal{V}_{0}\theta L^{3}\sqrt{\tau t N_{0} N_p}\left(l^{2}+n^{2}\right)},
\end{align}
where we used $M=\tau / t$, with $\tau$ the integration time and $t$ is the interaction time, which is bounded by the lifetime of the phonons. 
Similarly, one can estimate the precision for the the mode-mixing channel, where $A=8|\mathcal{M}_{l n}\mathcal{V}_{l,n}|t/\hbar$. However in this case $l,n$ are mode numbers of frequency modes whose difference matches $\Omega$.

Note that when the phonons interact with a periodic potential which encodes the parameter to be estimated, it has been found to be convenient to choose phonon modes that are in resonance with the frequency of the potential. Previous work shows that resonances producing parametric amplification lead to higher degrees of mode mixing and squeezing of phonons \cite{PhysRevLett.111.090504,Bruschi_2013,R_tzel_2018}. The periodic potentials can be due to  gravitational \cite{GWDetection,R_tzel_2018} or electromagnetic \cite{PhysRevLett.109.220401,bravo_analog_2015,PhysRevD.98.025011,pJuschitz} effects and lead to higher sensitivities in the detectors \cite{Accelerometer,GWDetectorFirst,GWDetectorFirst,Accelerometer}.

\emph{Application of frequency interferometry to gravitational wave detection}. In the case that the external potential corresponds to a monochromatic continuous gravitational wave of frequency $\Omega=\omega_l+\omega_n$, the potential is $\mathcal{V}_{\epsilon}(x,t) =  \mathcal{V}_0  x^2 \sin \Omega t$, with $\mathcal{V}_0=m\Omega^2/4$ (see \cite{maggiore2008gravitational,howl2018comment} and Appendix \ref{app:GW}). Consider, for example, a $^{7}{\mathrm{Li}}$ BEC with $N_0 = 4.4 \times 10^8$ atoms \cite{Straten2007} in a trap with dimensions $L= 0.005\unit{m}$ and $\alpha = 0.001$. We assume that the scattering length is tuned to $a = 99a_0$, with $a_0$ the Bohr radius. Taking into consideration two and three-body losses, we initially prepare the modes $n=260$ and $l = 258$  in a two-mode squeezed state with $N_p = 4000$ phonons and  a lifetime of $t=1\unit{ms}$. In this case, we obtain a sensitivity of $\Delta \epsilon = 4.3 \times 10^{-21}$ for the interferometry scheme at a gravitational wave of angular frequency $\Omega =25.3 \unit{kHz}$ and assuming $10$ independent detectors operating for $\tau = 1 \unit{yr}$. In contrast, the previous non-interferometric scheme would have provided a sensitivity of $\Delta \epsilon = 10^{-10}$. Higher sensitivities can be reached because frequency interferometry increases the number of phonons in the side modes by beam-splitting the phonons with the condensate, which can be achieved by, for example, modifying the tapping potential (see \cite{PhononEvap,HeterodyneBECs} and Appendix \ref{app:BEC}). For a detail derivation, constraints and examples on sensitivities, see Appendix \ref{app:GW}.  

There are no known astrophysical objects which are small and dense enough to emit at frequencies beyond $10^4 \unit{Hz}$ \cite{Aggarwal2020}. However, we consider sensitivities to higher frequencies because any discovery beyond this range would be produced by other yet-unknown sources of gravitational waves in the cosmos. This includes exotic objects such as primordial black holes or boson stars and cosmological events in the early Universe, such as phase transitions, preheating after inflation, oscillons, and cosmic strings \cite{Aggarwal2020,Caprini_2018}. Detecting gravitational waves at frequencies beyond the range of sensitivity of LIGO might eventually lead to the detection of dark matter. Interestingly, ultralight dark matter models for collective phenomena, as opposed to single scattering regimes, predict monochromatic long-lived dark matter waves at very high frequencies. The frequency of these wave is set by the dark matter mass and ranges from $10^{-8}$ to $10^{14}$ Hz \cite{darkmatter20180}. The range of the application of frequency interferometry to the detection of long-lived gravitational waves depends on the frequency of the modes that resonate with the wave. In the case that the modes are BEC phonons, this is given by the mode number, the speed of sound and the dimensions of the trap. 

Interestingly, Weber bars attempted to use resonances of frequency modes to detect gravitational waves. However, the large metallic devices were not cooled below $100\,m\mathrm{K}$. At these temperatures the vibrational modes are in the classical domain and cannot be prepared in quantum states. Ultracryogenic detectors could reach the standard quantum limit, but squeezing of the modes could not be produced in these systems \cite{Aguiar_2010}. The proposal in \cite{GWDetectorFirst} resembles a quantum version of a Weber bar where the atom-atom interactions in the BEC are Hamiltonian non-linearities which produce quantum excitations of phononic modes. In this case, a harmonic perturbation can produce squeezing via parametric amplification exploiting resonances between the potential perturbation and the quantum modes.

 \emph{Compatibility with General Relativity}. Quantum spatial atom interferometers are commonly described by non-relativistic quantum mechanics. The state evolution is then given by the Schr\"{o}dinger equation which is invariant under Galilean transformations. For this reason, the description is only compatible with the Newtonian approximation of gravity where the notion of time is absolute. Describing spatial interferometers beyond the Newtonian approximation is  non-trivial since the equations, as well as the inner products, must be Lorentz-invariant and conserve quantum probabilities. This condition can be consistently  satisfied for quantum fields but is problematic for individual particles \cite{BirrelandDavies}. An important advantage of frequency interferometry is that it can be applied in both Newtonian and General Relativistic regimes, in the latter using a theoretical description that is underpinned by quantum field theory in curved spacetime \cite{Edward_Bruschi_2014,PhysRevD.98.025011}. This enables the application of the scheme to study effects in  General Relativity and modified theories, including the estimation of spacetime parameters, and searches for dark energy and matter.

\emph{Summary.} We have introduced a scheme for interferometry that uses the frequency modes of a quantum field trapped in a localized potential. This setup does not require a large spatial extent, facilitating miniaturization. Since the modes occupy the same region in space at all times, it is possible to estimate parameters encoded in both phase shift and global channels acting on the modes. This includes channels that entangle the modes. As an example, we estimate the parameters of two-mode Gaussian channels using a three-mode scheme that implements an analogue in the frequency domain of the tritter operation introduced in the pumped-up SU(1,1) scheme \cite{SU11}. The tritter operation improves the scaling of the precision with the number of particles. Quantum frequency interferometry should be implementable in many systems, such as optical, hybrid atom-light, superconducting circuits, cold atoms and BECs  where the global channels could be generated using, for example, non-linear mediums and non-linear interactions. We show how the three-mode example can be implemented in a BEC, where the scheme exploits atom-atom interactions. These type of interactions produce undesired noise in other setups such as atom spatial interferometry. When the external potential that interacts with the BEC corresponds to an electromagnetic or a gravitational field, the scheme can be used to estimate the field parameters with a system trapped in a millimetre-scale trap. We show that this scheme can improve the precision in the detection of gravitational waves, enabling good sensitivities even in the case that squeezing is much smaller than assumed previously in \cite{GWDetectorFirst} and that the system suffers from short phononic lifetimes. 
\vspace{0.5cm}
\begin{acknowledgments}

We thank Paul Juschitz, Jan Kohlrus, Daniel Goldwater, Tupac Bravo, Daniel Hartley and Dennis R\"{a}tzel for useful discussions and comments. R.H. and I.F. would like to acknowledge that this project was made possible through the support of a donation by John Moussouris and the grant `Leaps in cosmology: gravitational wave detection with quantum systems' (No. 58745) from the John Templeton Foundation. R.H. would also like to acknowledge the support of the ID 61466 grant from the John Templeton Foundation, as part of the QISS project. The opinions expressed in this publication are those of the authors and do not necessarily reflect the views of the John Templeton Foundation.

\end{acknowledgments}

\appendix

\onecolumngrid

\newpage
\section{Symplectic matrices of  interferometry operations} \label{app:SympleticMatrices}

Here we provide the symplectic matrices, in the real $q,p$ representation, for the various processes involved in our considered active interferometry schemes. As in the main text, the initial state of the pump mode is assumed to be a coherent state, and so the displacement and covariance matrices of the full  input state to the interferometer are:
\begin{align}
\bs{d}_0 &= \ba{c} 2 Re (\alpha) \\ 2 Im (\alpha) \\ 0 \\ 0 \\0 \\0\ea = \sqrt{\overline{N}} \ba{c} 2 \cos \vartheta_0 \\ 2 \sin \vartheta_0 \\ 0 \\ 0 \\0 \\0\ea,\\
\bs{\sigma}_0 &= \bs{1}_6,
\end{align}
where we have written $\alpha \equiv \sqrt{\overline{N}} e^{i \vartheta_0}$, with $\overline{N}$ the total particle number, and $\bs{1}_6 := \mathrm{diag}(1,1,1,1,1,1)$  the identity matrix of which the first two rows and columns are for the pump mode, the next two rows and columns are for one of the side modes, and the final rows and columns are for the other side mode. 

The first stage of the interferometer is   the  two-mode squeezing operation that parametrically populates the side modes, which has the following  symplectic matrix (see e.g.\ \cite{CMF}): 
\begin{align} \label{eq:Ss}
\bs{S}_{s} &= \ba{cccccc} 1 & 0 & 0& 0 & 0 & 0 \\
0 & 1 & 0& 0 & 0 & 0 \\
0 & 0 & \cosh r & 0  & \sinh r \cos \vartheta_{sq} & \sinh r \sin \vartheta_{sq} \\
0 & 0 & 0 & \cosh r & \sinh r \sin \vartheta_{sq} & - \sinh r \cos \vartheta_{sq} \\
0 & 0 & \sinh r \cos \vartheta_{sq} & \sinh r \sin \vartheta_{sq} & \cosh r & 0 \\
0 & 0 & \sinh r \sin \vartheta_{sq} & - \sinh r \cos \vartheta_{sq} & 0 & \cosh r \ea,
\end{align}
where $r$ is the squeezing parameter and $\vartheta_{sq}$ is the squeezing phase. Here, the first two columns and rows are for the pump, the next two column and rows are for one of the side modes, and the last two columns and rows are for the other side mode.

The next stage is  a tritter between the side-modes and  the pump, which has the  symplectic matrix given in \eqref{eq:Str}. Following the tritter, there is the squeezing or mode-mixing channel, which are defined by the unitary transformations:
\begin{align} \label{eq:UsApp}
U(\xi)&= e^{\xi \hat{a}_1\da \hat{a}_2\da - \xi^{\ast} \hat{a}_1 \hat{a}_2}~~\mathrm{or}\\ \label{eq:UmApp}
U(\zeta)&= e^{\zeta \hat{a}_1\da \hat{a}_2 - \zeta^{\ast} \hat{a}_1 \hat{a}_2\da},
\end{align}  
where $\xi:= s e^{i \phi_{B}}$ and $\zeta := m e^{i \phi_{A}}$, with  $s \geq 0$, $m \geq 0$ and $\phi_A, \phi_B \in \mathbb{R}$. The symplectic matrices for these unitary evolutions are (see e.g.\ \cite{CMF}):
\begin{align} \label{eq:Ssc}
\bs{S}_{sc} &= \ba{ccc} \bs{1} & \bs{0} & \bs{0} \\
\bs{0} &  \bs{1} \cosh s &   \bs{R_{\phi_{B}}} \sinh s \\
\bs{0} &  \bs{R_{\phi_{B}}} \sinh s &   \bs{1} \cosh s \ea ,
\end{align}
and 
\begin{align} \label{eq:Smc}
\bs{S}_{mc} &= \ba{ccc} \bs{1} & \bs{0} & \bs{0} \\
\bs{0} & \bs{1} \cos  m  &   \bs{R_{ \phi_{A}}} \sin m \\
\bs{0} & -\bs{R^T_{ \phi_{A}}}\sin m  &  \bs{1} \cos  m \ea,
\end{align}
where:
\begin{align}
\bs{R_{ \phi_{B}}} &:= \ba{cc} \cos \phi_{ {B}} & \sin \phi_{ {B} } \\ \sin \phi_{ {B}} &  -\cos \phi_{ {B}}\ea,\\
\bs{R_{ \phi_{A}}} &:= \ba{cc} \cos \phi_{ {A}} & \sin \phi_{ {A} } \\ -\sin \phi_{ {A}} &  \cos \phi_{ {A}}\ea.
\end{align}
In contrast, the symplectic matrix for a unitary phase evolution  $\hat{U}(\phi) = \exp (-i \phi \hat{N}/2)$ would be the following (see e.g.\ \cite{CMF}):
\begin{align} \label{eq:Spc}
\bs{S}_{pc} = \ba{ccccccc} 1 & 0 & 0& 0& 0& 0 \\ 0 & 1 & 0& 0& 0& 0 \\ 0 & 0 & \cos \frac{\phi}{2} & \sin \frac{\phi}{2} & 0 & 0\\
0 & 0& -\sin \frac{\phi}{2} & \cos \frac{\phi}{2} & 0 & 0 \\
0 & 0 & 0 & 0 & \cos \frac{\phi}{2}& \sin \frac{\phi}{2} \\
0 & 0 & 0 & 0 & -\sin \frac{\phi}{2}  & \cos \frac{\phi}{2} \ea.
\end{align}

\section{Derivation of symplectic matrix of  tritter}\label{app:TritterMatrix}

The tritter used in the pumped-up SU(1,1) interferometry scheme  is generated by the following Hamiltonian \cite{PumpedUpSU11}:
\begin{align} \label{eq:Htritter}
H_{tr} = \frac{\hbar G}{\sqrt{2}} \Big[ e^{i \vartheta} \hat{a}_0^{\dagger} (\hat{a}_1 + \hat{a}_2) + e^{-i \vartheta} \hat{a}_0 (\hat{a}^{\dagger}_1 + \hat{a}^{\dagger}_2)  \Big],
\end{align}
which, in the Heisenberg picture, results in:
\begin{align}
\hat{a}_{1,2} (\theta) &= \hat{a}_{1,2} \cos^2 (\theta/2) - \hat{a}_{1,2} \sin^2 (\theta / 2) - \frac{i e^{- i\vartheta}}{\sqrt{2}} \hat{a}_0 \sin \theta,\\
\hat{a}_0 (\theta) &= \hat{a}_0 \cos \theta - \frac{ i e^{i \vartheta}}{\sqrt{2}} (\hat{a}_1 + \hat{a}_2) \sin \theta,
\end{align}
where $\theta := G t$ is the angle for an evolution time $t$,  and $\vartheta$ is the phase. We can write the above transformation in  matrix form:
\begin{align}
\bs{a}(\theta) :=\bs{A} \bs{a}
\end{align}
where:
\begin{align}
\bs{a} :=  \ba{c} \hat{a}_0 \\ \hat{a}^{\dagger}_0 \\ \hat{a}_1\\ \hat{a}_1^{\dagger}  \\ \hat{a}_2 \\ \hat{a}_2^{\dagger} \ea,
\end{align}
\begin{align}
\bs{A} := \ba{cccccc} \cos \theta & 0&  -\frac{i e^{i\vartheta}}{\sqrt{2}}  \sin \theta & 0 & -\frac{i e^{i\vartheta}}{\sqrt{2}}  \sin \theta & 0 \\
0 & \cos \theta&  0& \frac{i e^{-i\vartheta}}{\sqrt{2}}  \sin \theta & 0 & \frac{i e^{-i\vartheta}}{\sqrt{2}}  \sin \theta  \\
-\frac{i e^{-i\vartheta}}{\sqrt{2}} & 0 & \cos^2 (\theta/2) & 0 &- \sin^2 (\theta/2) &0\\
0 & -\frac{i e^{i\vartheta}}{\sqrt{2}} & 0 & \cos^2 (\theta/2) & 0 &- \sin^2 (\theta/2)\\
-\frac{i e^{-i\vartheta}}{\sqrt{2}} & 0 & 
- \sin^2 (\theta/2) & 0 &\cos^2 (\theta/2) &0\\
0 & -\frac{i e^{i\vartheta}}{\sqrt{2}} & 0 & 
\cos^2 (\theta/2) & 0  &\cos^2 (\theta/2)
\ea  .
\end{align}
We now move to the real symplectic $q,p$ representation:
\begin{align}
\bs{q}  = \bs{Q} \bs{a},
\end{align}
where:
\begin{align}
\bs{q} := \ba{c} q_0 \\ p_0 \\ q_1 \\ p_1 \\ q_2 \\ q_2 \ea, 
\end{align}
and:
\begin{align}
\bs{Q} := \ba{cccccc} 1 & 1 & 0 & 0 & 0 &0 \\
-i & i & 0 & 0 & 0 &0 \\
0 & 0 & 1 & 1 & 0 &0 \\
0 & 0 & -i & i & 0 &0\\
0 & 0 & 0 & 0 & 1 &1\\
0 & 0 & 0 & 0 & -i &i
\ea.
\end{align}
The symplectic representation of the tritter transformation is then:
\begin{align}
\bs{q}(\theta) = \bs{Q} \bs{a} (\theta) = \bs{Q} \bs{A} \bs{a} = \bs{Q} \bs{A} \bs{Q}^{-1}  \bs{q} := \bs{S}_{tr} \bs{q},
\end{align}
where:
\begin{align} \label{eq:Str}
\bs{S}_{tr} &= \bs{Q} \bs{A} \bs{Q}^{-1}\\
&=\ba{cccccc} \cos \theta & 0 & \frac{1}{\sqrt{2}} \sin \theta \sin \vartheta & \frac{1}{\sqrt{2}} \sin \theta \cos \vartheta & \frac{1}{\sqrt{2}} \sin \theta \sin \vartheta  & \frac{1}{\sqrt{2}} \sin \theta \cos \vartheta \\
0 & \cos \theta & - \frac{1}{\sqrt{2}} \sin \theta \cos \vartheta  & \frac{1}{\sqrt{2}} \sin \theta \sin \vartheta  & - \frac{1}{\sqrt{2}} \sin \theta \cos \vartheta & \frac{1}{\sqrt{2}} \sin \theta \sin \vartheta  \\
- \frac{1}{\sqrt{2}} \sin \theta \sin \vartheta  & \frac{1}{\sqrt{2}} \sin \theta \cos \vartheta & \cos^2 (\frac{\theta}{2}) & 0 & \frac{1}{2} (-1 + \cos \theta) & 0 \\
-\frac{1}{\sqrt{2}} \sin \theta \cos \vartheta & \frac{1}{\sqrt{2}} \sin \theta \sin \vartheta  & 0 & \cos^2 (\frac{\theta}{2}) & 0 & \frac{1}{2} (-1 + \cos \theta) \\ -\frac{1}{\sqrt{2}} \sin \theta \sin \vartheta   & \frac{1}{\sqrt{2}} \sin \theta \cos \vartheta & \frac{1}{2} (-1 + \cos \theta) & 0 & \cos^2 (\frac{\theta}{2}) & 0\\
-\frac{1}{\sqrt{2}} \sin \theta \cos \vartheta & -\frac{1}{\sqrt{2}} \sin \theta \sin \vartheta & 0 & \frac{1}{2} (-1 + \cos \theta) & 0 & \cos^2 (\frac{\theta}{2}) \ea .
\end{align}
Note that with a conventional two-way beam-splitter, $\theta = \pi/2$ would swap the modes. However, for the above tritter, $\theta = \pi/2$ would not completely swap the side modes and pump modes. This is responsible for $N$ appearing in the QFI expressions even when $\theta = \pi/2$.

\section{Quantum Fisher Information of Phase-Shift Channel}\label{app:PhaseShift}

If a conventional phase-shift channel were used instead of the squeezing or mode-mixing channels, the QFI is:
\begin{align} \nonumber
H = \frac{1}{4} \Big[ &\sin^2 (2 \theta) \sinh^2 r + 2 (1 + \cos^4 \theta)  \sinh^2 (2r)\\ \label{eq:HPFull} &+|\alpha_0|^2 \Big(4 \sin^4 \theta + \eta_p(r) \sin^2 2 \theta \Big)\Big],
\end{align}
where:
\begin{align}
\eta_p(r) &:= \sinh (2r) \cos \nu_{P} + \cosh (2r),\\
\nu_{P} &:= 2 (\vartheta - \vartheta_P) + \vartheta_{sq}.
\end{align}
This is obtained using the CMF by replacing $\bs{S}_{\epsilon}$ with the symplectic matrix for a phase-shift channel. It was also derived in \cite{PumpedUpSU11} using the Heisenberg picture rather than the CMF \footnote{Due to differing conventions on phase, the expression for the QFI \eqref{eq:HPFull} exactly  matches that in \cite{PumpedUpSU11} when $\vartheta_{sq} \rightarrow -\vartheta_{sq} - \pi/2$ in \eqref{eq:HPFull}.}.

\section{Turning points of Quantum Fisher Information}\label{app:TurningPoints}

As discussed in the main text, the QFI $H$ of the interferometer with a squeezing channel has three turning points when the optimum phase relations are chosen: $\vartheta_{sq} = \phi_B + \pi/2$ and $\vartheta = \vartheta_0 - \phi_B/2 + \pi/4$. These occur at: $\theta = 0$, $\theta = \pi/2$ and $\theta = \theta_t$, where $\theta_t = \cos^{-1} (z_t)/2$  with:
\begin{align} \label{eq:thetat}
\theta_t &:= \cos^{-1} (z_t)/2,\\
z_t &:= \frac{\csch^2 r (\sinh(2r)^2 - 2|\alpha_0|^2)}{4 |\alpha_0|^2(1+ \coth r) - 2 \cosh(2r)} \equiv \frac{N (N+4) - 2 \overline{N}}{N (2 \overline{N} - 3 N - 1) + 2(\overline{N} - N) \sqrt{N(N+2)}}.
\end{align}
The angle $\theta_t$ matches that  found in \cite{PumpedUpSU11} for a phase-shift channel. When $\overline{N}$ is large it can be approximated by  \cite{PumpedUpSU11}:
\begin{align}
\theta_t \approx \frac{1}{4} \pi + \frac{1}{2} \csc^{-1} (N + \sqrt{N(N+2)}).
\end{align}

For $\theta = 0$, $\theta = \pi/2$ and $\theta = \theta_t$, $H$ is:
\begin{align} \label{eq:Hsq0}
H(\theta = 0) &= \frac{1}{4}B^2 \Big(1+ \sin^2 (\vartheta_{sq} - \phi_B) \sinh^2 (2r)\Big) \equiv \frac{1}{4} \Big(1+ \sin^2 (\vartheta_{sq} - \phi_B) N^2\Big),\\
H(\theta = \frac{\pi}{2}) &= \frac{1}{4}B^2 \Big(1+ N_0 + \frac{1}{2} \sin^2 (\vartheta_{sq} - \phi_B) N^2\Big),\\\label{eq:Hsqt}
H(\theta = \theta_t) &= \frac{1}{32}B^2  \overline{N}e^{2r} (1+ \coth r) + \mathcal{O}(\overline{N}^0) \\&\xrightarrow[r \gg 1]{}  \frac{1}{8}B^2 \overline{N} N,
\end{align}
where, for the $\theta = \theta_t$ turning point, we have assumed that $\overline{N} \gg 1$ and taken the optimum phase relation $\vartheta =  \vartheta_0 + \vartheta_{sq}/2 - 2 \phi_B$.	

Analogous expressions to \eqref{eq:Hsq0}-\eqref{eq:Hsqt} for the squeezing case can be obtained for $H$ of the mode-mixing channel when $\theta = 0$, $\theta = \pi/2$ with $\phi_A = \pi/2$, and  $\theta = \pi/2$ with $\phi_A = 0$:
\begin{align} \label{eq:Hm0}
H(\theta = 0) &= \frac{1}{4}A^2 N^2,\\
H(\theta = \frac{\pi}{2}, \phi_A = \frac{\pi}{2}) &= \frac{1}{4}A^2 \Big( N_0 + \frac{1}{2} N^2\Big),\\
H(\theta = \frac{\pi}{2}, \phi_A =0 ) &= \frac{1}{4}A^2  (\overline{N}e^{2r} + N) + \mathcal{O}(\overline{N}^0) \\&\xrightarrow[r \gg 1]{}  \frac{1}{2} A^2 \overline{N} N,
\end{align}
where, for the last case we have assumed that $\overline{N} \gg 1$ and taken    $\vartheta =  \vartheta_0 - \vartheta_{sq}/2 +\pi/2$.


\section{Sensitivity}\label{app:Sensitivity}

In the main text we assume an intensity measurement. For the squeezing channel, and working with small $s$, we find the following expressions for this measurement scheme:
\begin{align}\nonumber
\braket{\hat{S}} =& \frac{1}{4} s^2 \Big(|\alpha_0|^2 \sin ^2(2 \theta ) \left(\sinh (2 r) \cos \nu_B +\cosh (2 r)\right)\\ \nonumber &+4\left(1+ \cos ^4\theta \right) \left(\sinh ^2(2 r) \sin ^2(\vartheta_{sq}-\phi_{B} )+1\right)\\&+\sin ^2(2 \theta ) \cosh ^2 r\Big),\\ \nonumber
\mathrm{Var}(\hat{S}) =& \frac{1}{4} s^2 \Big(|\alpha_0|^2 \sin ^2(2 \theta ) (\sinh (2 r) \cos \nu_B +\cosh (2 r))\\\nonumber&+8\left(1+ \cos ^4 \theta \right) \left(\sinh ^2(2 r) \sin ^2(\vartheta_{sq}-\phi_{B} )+1\right)\\&+\sin ^2(2 \theta ) \cosh ^2 r\Big),\\
F_0 =&  \frac{1}{16} B^2 \frac{\Big(\eta_4 \sin^2 (2 \theta) + \left(1+ \cos ^4 \theta \right) \eta_5 \Big)^2}{   \eta_4 \sin^2 (2 \theta) + 2\left(1+ \cos ^4 \theta \right) \eta_5},
\end{align}
where:
\begin{align}
\eta_4&:= |\alpha_0|^2 \eta_1(r) + \cosh^2 r,\\
\eta_5&:=\sinh^2 (2r) \eta_2 + 1.
\end{align}
We now use particle conservation to write $|\alpha_0(r)|^2 = |\alpha|^2 - 2 \sinh^2 r$ and $|\alpha|^2 = \overline{N}$ so that $|\alpha(r)|^2 = 1/ \varepsilon  - 2 \sinh^2 r$, where $\varepsilon := 1 /\overline{N}$, and take $\overline{N} \gg 1$,  to find:
\begin{align} \label{eq:F0s}
F_0 &= \frac{1}{16} B^2 \sin^2 (2 \theta) \eta_1(r) \overline{N} + \mathcal{O} (\overline{N}^{0})\\
&\approx \frac{1}{8}B^2 \sin^2 (2 \theta) N \overline{N}, 
\end{align}
where in the last line we have assumed that $N \gg 1$ and taken  the optimum phase relation  $\nu_B = 0$.  

If, on the other hand, the mode-mixing channel is chosen, then, for small $m$, we have:
\begin{align}\nonumber
\braket{\hat{S}} &=   m^2 \Big[ \sin^2 \theta ( \Phi_1 |\alpha_0| ^2  \eta_3(r) -  \Phi_2 \sinh ^2 r \\&+(\Phi_1-1) \sinh ^2(2 r)  )+2 \sinh ^2(2 r)\Big],\\\nonumber
\mathrm{Var}(\hat{S}) &= m^2 \Big[\Phi_1  \sin ^2\theta (|\alpha_0|^2 \eta_3(r)-\sinh ^2 r\\&+ 2  \sinh ^2(2 r))+2 (1 + \cos^2 \theta) \sinh ^2(2 r)\Big],
\end{align}
where
\begin{align}
\Phi_2 &:= \sin ^2(\theta )  \cos^2( \phi_{A} )-1.
\end{align}
Taking $\overline{N} \gg 1$, we find:
\begin{align} \label{eq:F0m}
F_0 &= \frac{1}{4} A^2 \sin^2 \theta~ \Phi_1 (\theta,\phi_A) \eta_3(r) \overline{N} +  \mathcal{O} (\overline{N}^{0})\\
&\approx \frac{1}{2 }A^2 \sin^2 \theta (1-\sin^2 \theta \sin^2 \phi_A)\overline{N} N , 
\end{align}
where in the last line we have assumed that $N \gg 1$ and taken  the optimum phase relation  $\nu_A = \pi$. 

\section{Full undepleted pump regime}\label{app:SmallTheta}

In the main text, we have assumed that the pump is relatively undepleted after the first active element (see also \cite{PumpedUpSU11}). If we want to further assume that the pump is also relatively undepleted after the tritter stage, then $\theta$ cannot be too large.  After the tritter stage, in general, the number of particles in the pump and side modes is the following:
\begin{align}
N_0(\theta) &= N_0 \cos^2 \theta + \frac{1}{2} N \sin^2 \theta,\\
N(\theta) &= N_0 \sin^2 \theta + \frac{1}{2} N (1 + \cos^2 \theta ).
\end{align}
Let us require that $N = \gamma N_0$ and $N(\theta) = \delta N_0 (\theta)$  where $\gamma \ll 1$, $\delta \ll 1$ and $\delta \geq \alpha$. Then $\theta$ must satisfy:
\begin{align} \label{eq:theta}
\theta \leq \frac{1}{2}  \arccos \Big(\frac{\delta \gamma + 2 \delta - 3 \gamma - 2}{\delta \gamma - 2 \delta + \gamma - 2}\Big).
\end{align} 
For example, taking $\delta = 0.1$,  we obtain:
\begin{align}
\theta \leq \frac{1}{2}  \arccos \Big(\frac{18 + 29 \gamma}{22 - 11 \gamma}\Big),
\end{align} 
which, in the limit $\gamma \rightarrow 0$, gives $\theta^2 \approx 0.0938$, and we note that $\theta^2 / \sin^2 \theta \approx 1.03$.

In the case of the squeezing channel, the QFI in this fully undepleted pump regime becomes:
\begin{align} \label{eq:Hsqsmalltheta}
H &\approx \frac{1}{4} B^2 \Big( 1 + N^2 + \theta^2 ( N_0 e^{2r} + N/2 - N^2) \Big)\\ \label{eq:pumpedQFIsq}
&\approx \frac{1}{2} B^2 \theta^2 N_0 N ,
\end{align}
where we have again used $\vartheta =  \vartheta_0 + \vartheta_{sq}/2 - 2 \phi_B$, and further  assumed that $ r \gg 1$ and $\overline{N} \gg 1$ in the last line. When considering a number-sum measurement scheme, the square-inverse of the sensitivity is:   
\begin{align} \label{eq:F0Squeezed}
F_0 \approx \frac{1}{2} B^2 \theta^2 N_0 N,
\end{align}
when  $r \gg 1$, and  $2 \vartheta = 2 \vartheta_0 + \vartheta_{sq} - 2 \phi_{B}$. This agrees with the QFI expression  above \eqref{eq:pumpedQFIsq}. The  number-sum measurement is, therefore, an optimum measurement scheme in these limits. 

For the pump to remain relatively
undepleted before the mode-mixing channel, the QFI becomes:
\begin{align}  \label{eq:Hmsmalltheta}
H &\approx \frac{1}{4} A^2 \Big(  N^2 + \theta^2 ( N_0 e^{2r} + N/2 - N^2) \Big)\\ \label{eq:pumpedQFIm}
&\approx \frac{1}{2} A^2 \theta^2 N_0 N ,
\end{align}
where we have  assumed that $ r \gg 1$ in the last line. This is similar to the QFI for the squeezing channel \eqref{eq:Hsqsmalltheta}, just with $B$ replaced by $A$. Equally, the square inverse of the sensitivity is:  
\begin{align} \label{eq:F0ModeMixed}
F_0 \approx \frac{1}{2} A^2 \theta^2 N_0 N,
\end{align}
with $2 \vartheta = 2 \vartheta_0 - \vartheta_{sq}$ and $r \gg 1$. 

\section{Heterodyne detection}\label{app:Heterodyne}

Rather than using a number-sum measurement, another possibility would be to use a  heterodyne measurement, for example, between the pump and the side modes.  Balanced homodyne detection for the side modes was considered in \cite{Li_2014} for a standard SU(1,1) interferometer and \cite{TruncatedSU11} for a `truncated' SU(1,1) experiment. In our considered heterodyne  case, at the measurement stage a balanced beam splitter could be applied between one of the side modes and the pump, and the difference of the number of particles in the two output parts of the final beam splitter could be considered: $\hat{S} = \hat{N}_1 - \hat{N}_2$. In the covariance matrix formalism we have:
\begin{align}
\braket{\hat{S}} &= \frac{1}{4} [ Tr(\sigma J_z) + d^T J_z d]\\
\mathrm{Var}(\hat{S}) &= \frac{1}{8} [Tr([\sigma J_z]^2 ) + 2 d^T J_z \sigma J_z d - 2n].
\end{align}
where:
\begin{align}
J_z = \ba{cc} \bs{1} & \bs{0} \\ \bs{0} & - \bs{1} \ea.
\end{align}
However, in order to measure the squeezing  parameter of the estimation channel, $\mathrm{Var}(\hat{S})$ or $\braket{\hat{S}^2}$ would need to be considered as the signal (see e.g.\ \cite{KnightBook}) and, therefore, the variance of this would be used in the error estimation.

\section{Implementation in a BEC}\label{app:BEC}

In this section, we provide detail on the implementation of the three-mode frequency interferometric scheme using phonons of a BEC. The Hamiltonian of a  Bose gas with a tapping potential $\mathcal{V}(\bs{r})$ is:
\begin{align} \label{eq:fullH3}
\hat{H} =  &\int_{V}  d\bs{r} \hat{\Psi}^{\dagger} \Big[-\frac{\hbar^2}{2m} \nabla^2 + \mathcal{V}(\bs{r}) \Big]\hat{\Psi}
+ \frac{1}{2} g \int_{V}  d\bs{r} \hat{\Psi}^{\dagger} \hat{\Psi}^{\dagger} \hat{\Psi} \hat{\Psi},
\end{align}
where $V$ is the volume.  We consider the {\it dilute regime}, where the contact potential only depends on two-body interactions and the coupling strength is given by $g = 4 \pi \hbar^2 a / m$, where $a$  isthe s-wave scattering length \cite{PitaevskiiBook,LandauLifshitzQM}. It is convenient to decompose the field operator $\hat{\Psi}^{\dagger}$ in terms of the annihilation operator for the ground state $\hat{a}_0$ and the operators for the $n$-th excited state $\hat{a}_n$:
\begin{align} \label{eq:PsiExpand}
\hat{\Psi} (\boldsymbol{r},t) &=  [\hat{\psi}_0(\bs{r}) + \hat{\psi}(\bs{r},t) ] e^{- i \mu(t) / \hbar},
\end{align}
where:
\begin{align}
\hat{\psi}_0(\bs{r}) &:= \phi_0(\bs{r}) \hat{a}_0,\\
\hat{\psi}(\bs{r},t) &:= \sum_{n \neq 0} \phi_{n} (\bs{r}) \hat{a}_{n}(t).
\end{align}
with the commutator $[\hat{a}_n,\hat{a}^{\dagger}_l]=\delta_{nl}$. We consider $\mu(t) = \mu t$ where $\mu$ is the chemical potential. In the case that the external potential vanishes, the Hamiltonian can be diagonalized by taking two steps. The first is assuming that the ground state is macroscopically occupied in a large coherent state. In this case the ground state operator can be replaced by $\hat{a}_0\approx\sqrt{N_0}$. This is called the {\it Bogoliubov approximation} and it is applicable to systems with a large number of particles in the low temperature regime ($T$ much smaller than the condensates' critical temperature). The second step involves applying a Bogoliubov transformation to $\hat{a}_n$, such that:
\begin{align} \label{eq:BogoTransP} 
\hat{\psi}(\bs{r},t) = \sum_n [ u_n (\bs{r}) \hat{b}_n e^{- i\omega_n t} + v^{\ast}_n (\bs{r}) \hat{b}\da_n e^{i \omega_n t} ],
\end{align}
where $[\hat{b}_n,\hat{b}^{\dagger}_l]=\delta_{nl}$. Neglecting trilinear and quartic terms in $\hat{b}_n,\hat{b}\da_n$ (since they have fewer factors of $\sqrt{N_0} \gg 1$), yields the diagonal Hamiltonian (see e.g.\ \cite{PitaevskiiBook}):
\begin{align} \label{eq:H0}
:\hat{H}: ~= \sum_n \hbar \omega_n \hat{b}\da_n \hat{b}_n.
\end{align}
Here $: :$ refers to normal ordering and $u_n,v_n$ are mode solutions to the Bogoliubov-de-Gennes equations \cite{PitaevskiiBook}:
\begin{align}
\hbar \omega_n u_n(\bs{r}) &= \Big[ -\frac{\hbar^2}{2m} \nabla^2 - \mu + 2 g N_0 |\phi_0|^2 \Big] u_n (\bs{r}) +  g N_0 \phi_0^2 v_n (\bs{r})\\  
-\hbar \omega_n v_n(\bs{r}) &= \Big[ -\frac{\hbar^2}{2m} \nabla^2 - \mu + 2 g N_0 |\phi_0|^2 \Big] v_n (\bs{r}) + g N_0 \phi^{\ast 2}_0 u_n (\bs{r}),
\end{align}
satisfying the orthonormal condition:
\begin{align}
\int d \bs{r} [ u_n^{\ast} (\bs{r}) u_l (\bs{r}) - v^{\ast}_n (\bs{r}) v_l (\bs{r}) ] = \delta_{nl}.
\end{align}
The ground state wave function $\phi_0$ satisfies the time-independent Gross-Pitaevskii equation:
\begin{align}
\Big[ -\frac{\hbar^2}{2m} \nabla^2 + \mathcal{V}(\bs{r})+  g N_0 |\phi_0|^2 \Big]\phi_0  = \mu \phi_0,
\end{align}
such that $\phi_0(t) := \phi_0 e^{-i \mu t / \hbar}$ satisfies the time-dependent version. The energy spectrum is given by:
\begin{align}
(\hbar \omega_n)^{2}=\left(c_{s} \hbar k_n\right)^{2}+\left(\hbar^{2} k^{2}_n / 2 m\right)^{2},
\end{align}
where $c_s := \sqrt{g \rho / m}$ is the speed of sound of the BEC and $\rho$ is the number density. In the {\it low energy limit} $\hbar \omega_n  \ll m c^2_s$, the dispersion law is linear and thus $\hat{b}\da_{n}$ and $\hat{b}_n$ create and annihilate phonons of the BEC. 

We now apply a small time-dependent potential $\epsilon \mathcal{V}_{\epsilon} (\bs{r},t)$ to the BEC where $\epsilon \ll 1$. This introduces a term $\epsilon \mathcal{V}_{\epsilon} \hat{\Psi}\da \hat{\Psi}$ to \eqref{eq:fullH3}, which, after applying \eqref{eq:PsiExpand} and \eqref{eq:BogoTransP}, provides an interaction Hamiltonian (see Appendix C of \cite{MassBEC} for a detailed derivation using the grand canonical Hamiltonian):
\begin{align} \label{eq:Hint}
\hat{H}_I (t) = \epsilon \int d \bs{r} \mathcal{V}_{\epsilon} (\bs{r},t) \Big[ |a_0|^2 |\phi_0|^2  &+ |a_0| \sum_{n} \Big(   \hat{b}_n e^{- i \vartheta (\bs{r}) } e^{-i\omega_n t}  +   \hat{b}\da_n e^{i \vartheta (\bs{r}) }  e^{i\omega_n t}\Big)\\
&+\sum_{n,m} [u_n^{\ast} (\bs{r}) u_m (\bs{r}) \hat{b}_n\da \hat{b}_m e^{i(\omega_m - \omega_n) t} + v_n (\bs{r})  v_m^{\ast}  (\bs{r}) \hat{b}_n \hat{b}_m\da e^{-i(\omega_m - \omega_n) t}] \\
&+\sum_{n,m} [u_n^{\ast}  (\bs{r}) v^{\ast}_m  (\bs{r}) \hat{b}_n\da \hat{b}\da_m e^{i(\omega_m + \omega_n) t} + u_n  (\bs{r}) v_m  (\bs{r}) \hat{b}_n \hat{b}_m e^{-i(\omega_m + \omega_n) t}]\Big],
\end{align}
where $\exp (i \vartheta(\bs{r})):= \phi_0 (\bs{r}) u^{\ast}_n (\bs{r}) + \phi^{\ast}_0 (\bs{r}) v^{\ast}_n (\bs{r})$.  

We now consider how the experimentalist can tailor a sequence of external potentials of the form $\mathcal{V}_{\epsilon}(\bs{r},t)$ with the appropriate resonance conditions to implement the unitaries $\hat{U}(\boldsymbol{\theta})$ and $\hat{U}(\boldsymbol{-\theta})$ in the phases (i) and (iii) of our frequency interferometric scheme. Frequency interfereometry can then be performed with a single BEC in a pumped-up SU(1,1) scheme. In this case, the condensate atoms act as the pump, and two phonon modes can be used as the side modes.

\subsection{Tritter}

The Hamiltonian for a tritter is given by \eqref{eq:Htritter} where here we treat $\hat{a}_0$ as the annihilation operator for the condensate, and $\hat{a}_{n \neq 0}$ as the annihilation operator for the phonon modes, which we denoted as $\hat{b}_{n \neq 0}$ above. Since the condensate must be more populated than the phonon modes before and after the tritter for our description of the BEC used above to still hold, we can apply the Bogoliubov approximation and drop the hat on $\hat{a}_0$, leaving us with:
\begin{align}
H_{tr} = \frac{\hbar G}{\sqrt{2}} |a_0| \Big[ e^{i \vartheta} (\hat{b}_m + \hat{b}_n) + e^{-i \vartheta} \hat{a}_0 (\hat{b}^{\dagger}_m + \hat{b}^{\dagger}_n)  \Big].
\end{align}
This can be picked out from \eqref{eq:Hint} by choosing an oscillating potential of the form $V(t) = \epsilon V_0 \cos (\Omega t) \cos(\Omega' t)$, where $\Omega := \omega_m + \omega_n$ and $\Omega' := \omega_n - \omega_n$, and assuming that $\vartheta_n(\bs{r}) \approx \vartheta_m(\bs{r})$, which could be achieved, for example, by choosing modes with equal and opposite momenta in a uniform BEC with periodic boundary conditions \cite{PitaevskiiBook}.

\subsection{Two-mode squeezing}

To create a two-mode squeezed state of phonons, we require a Hamiltonian of the form (see e.g.\ \cite{CMF}):
\begin{align}
\hat{H} = s [ e^{i \vartheta_{sq} } \hat{b}\da_m \hat{b}\da_n + e^{-i \vartheta_{sq} } \hat{b}_m \hat{b}_n] .
\end{align}
This can be obtained from \eqref{eq:Hint} by choosing an oscillating potential to pick out these particular terms on resonance \cite{MassBEC}. For example $V_{\epsilon}(t) = \epsilon V_{0}  \sin \Omega t $ would achieve this where $\Omega := \omega_m + \omega_n$ and $V_{0}$ is a constant amplitude.

\section{Application to the detection of gravitational waves}\label{app:GW}

As in the previous appendix, we assume a Bose gas operating in the dilute regime. However, we now take  a box potential with $\mathcal{V}(\bs{r}) = 0$ and $V = L \mathcal{A}$, where $L$  is the length and $\mathcal{A}$ is the cross-section. We assume that $L$ is much greater than the dimensions of $\mathcal{A}$ and restrict the analysis to modes with vanishing transversal wave numbers, i.e. we only consider the direction along $L$, which we label $x$. The mode solutions, which fulfill von Neumann boundary conditions since $\rho$ vanishes at the potential walls, are given by \cite{PitaevskiiBook,MassBEC}:
\begin{align}
u_n(\bs{r})&= u_n\sqrt{\frac{2}{L\mathcal{A}}} \cos(k_n (x +L/2)),\\
v_n(\bs{r})&= v_n \sqrt{\frac{2}{L\mathcal{A}}} \cos(k_n (x +L/2)),
\end{align}
where
\begin{align}
u_n, v_n = \frac{\sqrt{\hbar^2 k_n^2 / 2m} \pm \sqrt{2 m c^2_s + \hbar^2 k_n^2 / 2m}}{2 \sqrt{\hbar \omega_n}}
\end{align} 
and $k_n = n \pi / L = \omega_n/c_s$.  Taking the time-dependent potential $\mathcal{V}_{\epsilon} (\bs{r},t) = \mathcal{V}_{\epsilon} (x,t)= \mathcal{V}_{\epsilon} (x) \sin \Omega t$, writing $\sin \Omega t \equiv 1 / (2 i) (e^{i \Omega t} - e^{-i \Omega t})$, and applying the rotating-wave approximation, yields \cite{MassBEC}:
\begin{align}
\hat{H}_{I}(t)=& \sum_{n} M_{0 n}\left(\hat{b}_{n} \mathrm{e}^{-\mathrm{i}\left(\omega_{n}-\Omega\right) t}-\hat{b}_{n}^{\dagger} \mathrm{e}^{\mathrm{i}\left(\omega_{n}-\Omega\right) t}\right)+\sum_{l, n} M_{l n}\left(\hat{b}_{l} \hat{b}_{n} \mathrm{e}^{-\mathrm{i}\left(\omega_{n}+\omega_{l}-\Omega\right) t}-\hat{b}_{l}^{\dagger} \hat{b}_{n}^{\dagger} \mathrm{e}^{\mathrm{i}\left(\omega_{n}+\omega_{l}-\Omega\right) t}\right) \\ &-\sum_{l>n}\left(A_{l n}\left(\hat{b}_{l}^{\dagger} \hat{b}_{n} \mathrm{e}^{\mathrm{i}\left(\omega_{l}-\omega_{n}-\Omega\right) t}-\hat{b}_{n}^{\dagger} \hat{b}_{l} \mathrm{e}^{-\mathrm{i}\left(\omega_{l}-\omega_{n}-\Omega\right) t}\right)+B_{l n}\left(\hat{b}_{l}^{\dagger} \hat{b}_{n} \mathrm{e}^{\mathrm{i}\left(\omega_{l}-\omega_{n}-\Omega\right) t}-\hat{b}_{n}^{\dagger} \hat{b}_{l} \mathrm{e}^{-\mathrm{i}\left(\omega_{l}-\omega_{n}-\Omega\right) t}\right)\right),
\end{align}
with:
\begin{align}
M_{0 n} &\approx -\mathrm{i}  L^{3 / 2} \sqrt{\frac{N_{0} \zeta}{(\sqrt{2} n \pi)^{3}}}\left(\left(1+(-1)^{n}\right) \mathcal{V_{0}}\right)\\
M_{l n} &\approx-A_{l n} \approx-B_{l n} \approx \mathrm{i}  \frac{L^{3}\left(l^{2}+n^{2}\right)}{2 \sqrt{2 n l}\left(l^{2}-n^{2}\right)^{2} \pi^{3} \zeta} \mathcal{V}_{l,n} \text { for } l \neq n,\\
M_{n n} &\approx-A_{n n} \approx-B_{n n} \approx \mathrm{i}  \frac{L^{3} \mathcal{V}_{0} }{8 \sqrt{2} n^{3} \pi^{3} \zeta}
\end{align}
and $\zeta := \hbar / (\sqrt{2} m c_s)$ is the healing length.
If the external potential is linear $\mathcal{V}_{\epsilon}(x)= \mathcal{V}_{0} x$, where $\mathcal{V}_{0}$ is a constant, then the coefficients are given by $\mathcal{V}_{l,n}=-\left(1-(-1)^{l+n}\right) (\mathcal{V}_{0}/L)$ and, if it is quadratic $\mathcal{V}_{\epsilon}(x)=\mathcal{V}_{0}x^2$, then $\mathcal{V}_{l,n}=\left(1+(-1)^{l+n}\right) \mathcal{V}_{0}$ \cite{R_tzel_2018}.

From inspection of the interaction Hamiltonian, we see that the resonance condition $\Omega = \omega_n$ generates a displacement of the mode $n$ which creates a (classical) coherent state. Two-mode squeezing (parametric amplification) is produced through the resonance condition $\Omega = \omega_n + \omega_m$,  and $\Omega = \omega_n - \omega_m$ leads to mode-mixing (frequency conversion).  In the two-mode squeezing case the squeezing parameter is $s=\frac{\epsilon B}{4}=2 \epsilon |\mathcal{M}_{l n}\mathcal{V}_{l,n}|t/\hbar$ where 
\begin{align}
\mathcal{M}_{l n} & \approx \mathrm{i}  \frac{L^{3}\left(l^{2}+n^{2}\right)}{2 \sqrt{2 n l}\left(l^{2}-n^{2}\right)^{2} \pi^{3} \zeta},
\end{align}
with  $l,n$ ($l \neq n$)  the mode numbers of the two frequency modes whose sum is resonant with $\Omega$.  For the mode-mixing channel $A=8|\mathcal{M}_{l n}\mathcal{V}_{l,n}|t/\hbar$, however, $l,n$ are mode numbers of frequency modes whose difference sum is resonant with $\Omega$. 

Using the results also from the previous appendix, we can perform frequency interfereometry with a single BEC in a pumped-up SU(1,1) scheme, with the condensate atom acting as the pump, and two phonon modes as the  side modes. Since the number of condensate atoms $N_0$  must be much larger than the number of phonons $N_p$, with the implementation of a number-sum measurement scheme, the sensitivity of the interferometer to a two-mode squeezing channel:  
\begin{align}
    \Delta \epsilon &= \frac{1}{\sqrt{M F_0}}\\
    &\approx \frac{\sqrt{2}}{B \theta \sqrt{M N_0 N_p N_d}}\\
    &= \frac{\sqrt{2}\hbar}{8 |\mathcal{M}_{l n}\mathcal{V}_{l,n} |t\theta\sqrt{M N_{0} N_p N_d}}\\ \label{eq:DeltaEpsSq}
    &=\frac{\hbar^2 \pi^{3} { \sqrt{2 n l}\left(l-n\right)^{2}}}{4 m c_s|\mathcal{V}_{l,n} |\theta L^{3}\sqrt{\tau t N_{0} N_p N_d}{\left(l^{2}+n^{2}\right)}},
\end{align}
where $M$ is the number of repetitions of estimation and $\theta$ must satisfy \eqref{eq:theta}. In the second line we have assumed that $\theta^2 N_0 N_p \gg N_P^2$, and in the last line we have assumed that $M = \tau / t$ where $\tau$ is the total time of the full estimation procedure, with $t$ then being the time it takes for each individual estimation (which we take to be the phonon lifetime). $N_d$ is the number of independent detectors.

We now consider a gravitational wave interacting with a BEC. Gravitational waves are often considered from the perspective of the  transverse-traceless (TT) frame. However, this is not the frame of a BEC experimentalist, which can lead to intuitive effects  that must be considered carefully, such as the fact that a rigid cavity oscillates in this frame. Instead, we consider the BEC from the proper detector frame \cite{maggiore2008gravitational,howl2018comment}, which is the frame closest to a BEC experimentalist (for example, rigid cavities do not appear to change length in this frame). In this case, the effect of a monochromatic gravitational wave of frequency $\Omega$ on a quasi-one dimensional BEC is to introduce a time-dependent quadratic potential of the form   $\mathcal{V}_{\epsilon} (x,t)=  m x^2 \Omega^2 \sin \Omega t / 4$. Comparing to the above, the gravitational wave will create parametric-amplification in the BEC, leading to squeezing of phonons \cite{howl2018comment}. Using \eqref{eq:DeltaEpsSq}, and taking $\Omega =\omega_l+\omega_n$ with $l+n$ even, we find:
\begin{align}
\Delta \epsilon &=\frac{\hbar^2 \pi {\sqrt{2 n l}\left(l-n\right)^{2}}}{m^2 c^3_s \theta L\sqrt{\tau t N_{0} N_p N_d}{\left(l^{2}+n^{2}\right)}}\\ &=\frac{m \pi }{4 \sqrt{2}\hbar}\frac{\alpha^{3}}{\theta N^2_0 \sqrt{N_p N_d\tau t }}\sqrt{\frac{L^7}{a^3}}\frac{\sqrt{ n l}\left(l-n\right)^{2}}{\left(l^{2}+n^{2}\right)}\nonumber,
\end{align}
where we have used $c_s = \sqrt{g \rho / m}$ with $\rho = N_0 / V$, defined $\mathcal{A} := \pi R^2$ with $R =: \alpha L$, and $N_d$ is the number of independent detectors. This improves on  sensitivity of \cite{GWDetectorFirst}, which, written in fundamental experimental parameters, would be:
\begin{align}
\Delta \epsilon &\approx \frac{2}{\sqrt{M \omega_n\omega_l } N_p t}\\
&= \frac{m}{\sqrt{\pi}\hbar}\frac{\alpha }{\sqrt{N_0 N_d\tau t} N_p} \sqrt{\frac{L^5}{a}}\frac{1}{\sqrt{nl}}
\nonumber
\end{align}
Note that the sensitivity reported in \cite{GWDetectorFirst} is given in terms of $\Delta \epsilon/\sqrt{\Omega}$. However, we use $\Delta \epsilon$ since it is a figure of merit which is more convenient for resonant detectors. The sensitivity reported in \cite{GWDetectorFirst} also has a different dependence on $N_0$. This is because a factor $1/\sqrt{N_0}$ was added assuming that the sensitivity scales with the number of atoms detected in the experiment. In this work, the dependence on $N_0$ is not assumed, it is obtained from first principles calculations. 

Taking, for example,  a $^{7}{\mathrm{Li}}$ BEC with $N_0 = 4.4 \times 10^8$, $N_p = 4000$, $a = 99a_0$ ($a_0$ is the Bohr radius), $L= 0.005\unit{m}$, $\alpha = 0.001$, and $t=1\unit{ms}$, and resonating with modes $n=260$ and $l = 258$, we obtain a sensitivity of $\Delta \epsilon = 4.3 \times 10^{-21}$ for the interferometry scheme at a gravitational wave of angular frequency $\Omega =25.3 \unit{kHz}$ and assuming $10$ independent detectors operating for $\tau = 1 \unit{yr}$. In contrast, the previous non-interferometric scheme would have provided a sensitivity of $\Delta \epsilon = 10^{-10}$. In deriving these results we have been careful to satisfy various theoretical and experimental constraints. For example, we have made sure that the chosen modes are phononic modes ($\hbar \omega_{n,l} \ll m c^2_s$), that we are operating in the dilute regime ($n a^3 \ll 1$), and that \eqref{eq:theta} is satisfied. We have also chosen a phonon lifetime that is much smaller than the expected BEC and phononic lifetimes from three-body and two-body (Beliaev and Landau) decay processes for a three-dimensional BEC \cite{Howl_2017}. However, it should be noted that three-body and two-body decay lifetimes would be expected to very much set an upper bound for the BEC of the detector since it is assumed to operate  in the quasi one-dimensional or pure one-dimensional regimes where decay processes are heavily suppressed \cite{Howl_2017,dennis2021}. In  \cite{qweber}, a comprehensive review of the gravitational wave detector will be provided with a detailed account of the sensitivities that can be achieved using frequency interferometry, and with a range of constrained experimental parameters.  Furthermore, in  \cite{pJuschitz}, we have investigated the amount of squeezing that can be generated for phonons by periodically shaking the trap as proposed above, finding that high levels of squeezing are theoretically possible. The results in \cite{pJuschitz} take into account Landau and Beliaev damping as well as three-body recombination. Future work will also consider how the  bandwidth of the detector could be tuned in practice and the corresponding relevant figure of merit.
 
In the tables below we show further examples of the precision reached by the phononic gravitational wave detector using  $^{7}{\mathrm{Li}}$ (in the $|F = 1, mF = 1\rangle$  hyperfine state)  and $^{87}{\mathrm{Rb}}$ BECs. The frequency interferometry (FI) scheme improves by many orders of magnitude the sensitivities given by \cite{GWDetectorFirst}. The integration time $\tau$ is one year, the tritter angle is $\theta=0.31$ and we considered $N_d=10$ independent detectors. For $^{7}{\mathrm{Li}}$, the three-body loss rate is calculated using the universal theory of Efimov physics \cite{BRAATEN2006259,PhysRevLett.83.1751}, which have been shown to match experimental data very well \cite{Pollack1683}. For $^{87}{\mathrm{Rb}}$, we assume a three-body rate coefficient of $4 \times 10^{-30} \unit{cm^{6}\, s^{-1}}$ \cite{PhysRevA.53.916}.

\begin{center}
    $^{7}${\bf Lithium BEC}
\end{center}

\begin{center}
\begin{tabular}{ |c| c| c| c| c| c| c| c| c| c| c| c| c| c| c| c| c|}
  \hline
 {\bf Parameter}& $L$ & $\alpha$ & $N_0$ & $N_p$ & $l$ & $n$& $t$ & $a/a_0$ & $n_0$&$\Omega$&$c_s$  &$\Delta\epsilon$& $\Delta\epsilon/\sqrt{\Omega}$\\
  \hline
{\bf Units}  &$\unit{mm}$&  &  & & &  & $s$ &  & $cm^{-3}$& kHz&$m/s$&& $\unit{Hz^{-1/2}}$\\
 \hline
 \hline
 \hline

 \hline
 \cite{GWDetectorFirst} scheme  & $4$ & 0.01 & $4.4\times10^9$ & $2500$ & 480 & 478 & 0.02 & 119 & $2.2\times10^{14}$ &$28.4$ &  0.04& $4.1\times10^{-11}$&$2.4\times10^{-13}$\\
 \hline 
 FI scheme  & $4$ & 0.01 & $4.4\times10^9$ & $2500$ & 480 & 478 & 0.02 & 119 & $2.2\times10^{14}$ &$28.4$ &  0.04 & $2.9\times10^{-21}$&\\
 \hline
 \cite{GWDetectorFirst} scheme  & 5 & 0.001 & $4.4\times 10^8$ & $4000$ & 260 & 258 & 0.001 & 99 & $1.1\times10^{15}$ &25.3 &  0.08& $1.0\times10^{-10}$&$6.4\times10^{-13}$\\
 \hline 
 FI scheme  & 5 & 0.001 & $4.4\times 10^8$ & $4000$ & 260 & 258 & 0.001 & 99 & $1.1\times10^{15}$ &$25.3$ &  0.08 & $4.3\times10^{-21}$&\\
 \hline
 \cite{GWDetectorFirst} scheme  & 4 & 0.01 & $4.4\times10^9$ & $4.4\times10^7$ & 480 & 478 & 0.04 & 99 & $2.2\times10^{14}$ &$25.9$ &  0.03& $1.8\times10^{-15}$&$1.1\times10^{-17}$\\
 \hline 
 FI scheme  & 4 & 0.01 & $4.4\times10^9$ & $4.4\times10^7$ & 480 & 478 & 0.03 & 99& $2.2\times10^{14}$ &$25.9$ &  0.08& $2.0\times10^{-23}$&\\
 \hline

\end{tabular}
\end{center}

\begin{center}
    $^{87}${\bf Rubidium BEC}
\end{center}

\begin{center}
\begin{tabular}{ |c| c| c| c| c| c| c| c| c| c| c| c| c| c| c| c| c|}
  \hline
 {\bf Parameter}& $L$ & $\alpha$ & $N_0$ & $N_p$ & $l$ & $n$& $t$ & $a/a_0$ & $n_0$&$\Omega$&$c_s$  &$\Delta\epsilon$& $\Delta\epsilon/\sqrt{\Omega}$\\
  \hline
{\bf Units}  & $\unit{mm}$ &  &  & & &  & $s$ &  & $cm^{-3}$& kHz&$m/s$&& $\unit{Hz^{-1/2}}$\\
 \hline
 \hline
 \hline

 \hline
 \cite{GWDetectorFirst} scheme  & $6$ & 0.0005 & $4.4\times10^8$ & $4400$ & 480 & 478 & 0.002 & 109.6 & $2.6\times10^{15}$ &$5.3$ &  0.01& $3.8\times10^{-10}$&$5.3\times10^{-12}$\\
 \hline 
 FI scheme  & $6$ & 0.0005 & $4.4\times10^8$ & $4400$ & 480 & 478 & 0.002 & 109.6 & $2.6\times10^{15}$ &$5.3$ &  0.01 & $4.5\times10^{-21}$&\\
 \hline
 \cite{GWDetectorFirst} scheme  & 4 & 0.003 & $4.4\times 10^9$ & $4.4\times 10^7$& 1680 & 1678 & 0.003 & 109.6 & $2.4\times10^{15}$ &25.65 &  0.01& $7\times10^{-15}$&$4.4\times10^{-17}$\\
 \hline 
 FI scheme  & 4 & 0.003 & $4.4\times 10^9$ & $4.4\times 10^7$ & 1680 & 1678 & 0.003 & 109.6 & $2.4\times10^{15}$ &$25.65$ &  0.01 & $6.3\times10^{-24}$&\\
 \hline

\end{tabular}
\end{center}
\bibliography{references3}
\bibliographystyle{apsrev4-1}

\end{document}